\RequirePackage{fix-cm}
\documentclass[smallextended]{svjour3}
\smartqed  

\usepackage{natbib}
\usepackage{cite}
\usepackage{url}
\usepackage{xspace}
\usepackage{amssymb}
\usepackage{listings}
\usepackage{tikz}
\usepackage{pifont}
\usepackage{tabularx,booktabs}
\usepackage{multirow}
\usepackage{pgfplots}
\usepackage[linkcolor={}]{hyperref}
\usepackage{pgfplotstable}
\usepackage{tcolorbox}
\usepackage{adjustbox}
\usepackage{wrapfig}
\usepackage{arydshln}
\usepackage{diagbox}
\usepackage{xcolor}
\usepackage{color, colortbl}
\usepackage{xifthen}
\usepackage{etoolbox}

\usepackage{amsmath}
\usepackage{array}
\usepackage{fixltx2e}
\PassOptionsToPackage{hyphens}{url}
\usepackage{booktabs}
\usepackage{xcolor}
\usepackage{amsmath}
\usepackage{algorithm}
\usepackage[noend]{algpseudocode}
\usepackage{pifont}
\usepackage{tabularx,booktabs}
\usepackage{flushend}
\usepackage{keyval}
\usepackage{subcaption}

\usepackage{enumitem}
\usepackage[T1]{fontenc}
\usepackage{pgfplots}
\usepackage{soul}
\usepackage{changepage}
\usepackage{blindtext}
\usepackage{hyperref}
\usepackage{balance}

\newcommand{\roundbox}[1]{
	\smallskip
	\noindent
	\begin{tikzpicture}
		\node[draw=black, rectangle, rounded corners](box){
			\begin{minipage}{\columnwidth}
				#1
			\end{minipage}
		};
	\end{tikzpicture}
}

\newcommand{\RQ}[1]{\textbf{RQ#1}\xspace} 
\newcommand{\requirement}[1]{\textbf{R#1}\xspace} 

\definecolor{darkred}{rgb}{0.75,0,0}
\definecolor{darkblue}{rgb}{0,0,0.75}
\definecolor{darkgreen}{rgb}{0,0.75,0}

\newcommand{\hide}[1]{}

\newcommand{\srm}{\textsc{SM}\xspace}
\newcommand{\fluentql}{\textsc{\textit{fluent}TQL}\xspace}
\newcommand{\codeql}{\textsc{CodeQL}\xspace}
\def\Yields{\Downarrow} 

\newcommand{\pie}[1]{%
	\begin{tikzpicture}
		\draw (0,0) circle (1ex);\fill (1ex,0) arc (0:#1:1ex) -- (0,0) -- cycle;
	\end{tikzpicture}%
}

\definecolor{pblue}{rgb}{0.13,0.13,1}
\definecolor{pgreen}{rgb}{0,0.5,0}
\definecolor{pred}{rgb}{0.9,0,0}
\definecolor{pgrey}{rgb}{0.46,0.45,0.48}

\usepackage{listings}
\lstdefinelanguage{Xtext}{
	morekeywords={grammar, with, hidden, generate, as, import, returns, current, terminal, enum},
	keywordstyle=[2]{\textbf},
	morecomment=[l]{//}, 
	morecomment=[s]{/*}{*/}, 
	morestring=[b]",
	tabsize=4}

\newcommand{\TAS}{\textsc{TAS}\xspace}

\newcommand{\outvalue}[1]{#1^\textsc{out}\xspace}
\newcommand{\invalue}[1]{#1^\textsc{in}\xspace}
\newcommand{\syntax}[1]{\texttt{#1}\xspace}
\newcommand{\Sanitizers}{\textit{Sanitizers}\xspace}
\newcommand{\Sources}{\textit{Sources}\xspace}
\newcommand{\Sinks}{\textit{Sinks}\xspace}
\newcommand{\RPropagators}{\textit{RequiredPropagators}\xspace}
\newcommand{\linenumber}[1]{\ref{#1}}
\newcommand{\listref}[1]{Listing~\ref{#1}}

\newcommand*{\affaddr}[1]{#1} 
\newcommand*{\affmark}[1][*]{\textsuperscript{#1}}

\begin{document}


\title{Fluently specifying taint-flow queries with \fluentql
}
\titlerunning{\fluentql} 

\author{Goran Piskachev \protect\affmark[1]\affmark[4]
	\and Johannes Sp{\"a}th \protect\affmark[3]
	\and Ingo Budde\protect\affmark[1]
	\and Eric Bodden \protect\affmark[1]\protect\affmark[2]
}

\authorrunning{Piskachev, et. al} 

\institute{Goran Piskachev \at
	\email{goran.piskachev@iem.fraunhofer.de}
	\and
	\affaddr{\affmark[1]Fraunhofer IEM, Paderborn, Germany}\\
	\affaddr{\affmark[2]Department of Computer Science, Paderborn University, Paderborn, Germany}\\
	\affaddr{\affmark[3]CodeShield GmbH}\\
	\affaddr{\affmark[4]Corresponding author}\\
}

\date{}

\maketitle

\begin{abstract}
	Previous work has shown that taint analyses are only useful if correctly customized to the context in which they are used. Existing domain-specific languages (DSLs) allow such customization through the definition of deny-listing data-flow rules that describe potentially vulnerable or malicious taint-flows. These languages, however, are designed primarily for security experts who are expected to be knowledgeable in taint analysis. Software developers, however, consider these languages to be complex. 
	
	This paper thus presents \fluentql, a query specification language particularly for taint-flow. \fluentql is internal Java DSL and uses a fluent-interface design. \fluentql queries can express various  taint-style vulnerability types, e.g.\ injections, cross-site scripting or path traversal. This paper describes \fluentql's abstract and concrete syntax and defines its runtime semantics. The semantics are independent of any underlying analysis and allows evaluation of \fluentql queries by a variety of taint analyses. Instantiations  of \fluentql, on top of two taint analysis solvers, Boomerang and FlowDroid, show and validate \fluentql expressiveness. 
	
	Based on existing examples from the literature, we have used \fluentql to implement queries for 11 popular security vulnerability types in Java. Using our SQL injection specification, the Boomerang-based taint analysis found all 17 known taint-flows in the OWASP WebGoat application, whereas with FlowDroid 13 taint-flows were found. Similarly, in a vulnerable version of the Java Spring PetClinic application, the Boomerang-based taint analysis found all seven expected taint-flows. In seven real-world Android apps with 25 expected malicious taint-flows, 18 taint-flows were detected. In a user study with 26 software developers, \fluentql reached a high usability score. In comparison to \codeql, the state-of-the-art DSL by Semmle/GitHub, participants found \fluentql more usable and with it they were able to specify taint analysis queries in shorter time. 
\end{abstract}

\begin{keywords}
	- taint analysis, program analysis, domain-specific language, user study, usability.
\end{keywords}


\section{Introduction}\label{sec-1-intro}

Over the past decade, static and dynamic taint analyses have gained significant traction both in industry and academia~\citep{popl19spds, soap2018secret, ultrascalable, flowdroid}. 
This is due to the fact that---in principle---most types of security vulnerabilities on the code level, e.g.\ 17 of the 25 vulnerabilities types of SANS-25~\citep{top25}, can be detected via taint analysis~\citep{swan}. 
Similarly, the OWASP top 10 list~\citep{top10} comprises 6 taint-style vulnerability types. 

Taint analysis tracks sensitive data from \emph{sources}, which are typically method calls to application programming interfaces (APIs), to program statements performing security-relevant actions, known as \emph{sinks}.
To soundly and precisely detect security vulnerabilities in a given software development project, any taint analysis, whether static or dynamic, requires configuration. Particularly, the sources and sinks must be configured regarding the libraries and frameworks the project uses~\citep{susi}.
Additionally, due to a lack of scalability, static analyses frequently are unable to analyze \emph{all} the software's code and must instead be configured to cut corners.

Some existing static analysis tools from academia~\citep{crySL, pql, pidgin} as well as from industry~\citep{checkmarx,fortify,codesonar,lgtm} provide a DSL to configure their analyses.
However, all of the existing DSLs are designed to be used by static analysis experts and not by software developers---despite the fact that developers are usually the ones who best know how the project under analysis is structured.
This was also confirmed in a recent research project with several industry partners~\citep{secucheck}, in which the authors conducted interviews with software developers that have used various commercial and non-commercial static analysis tools. Eight of the nine interviewees find the configuration options for taint analysis of the tools to be too complex. In another recent study among developers, the authors discovered that for 47.1\% of the participants, there is a dedicated team to configure the used static analysis tools, 36.8\% configure their analysis tools themselves, and 16.2\% run on default settings~\citep{tse20-2study}. Moreover, the existing DSLs require expertise in static code analysis which many developers do not have. We are not aware of any excising study that evaluates the usability of the DSLs used for configuring the tools. 

While much effort has been spent on automatically proposing relevant sources, sanitizers, and sinks \citep{swan, susi, sas} or inference of taint-flows (source-sanitizer-sink paths) \citep{merlin, inferencebigcode, inferring}, in practice taint analyses still require substantial manual specification effort. 

To address this shortcoming, this paper presents a new domain-specific language called \fluentql. \fluentql is designed for software developers---not static or dynamic analysis experts---and allows the specification of taint-flow queries. Compared to existing DSLs, the abstraction level of \fluentql is specific to taint analysis and contains only concepts that allow software developers to easily create or modify taint-flow queries. In result, \fluentql queries can be evaluated by virtually any existing taint analysis. This sets the language apart from previous more generic code-query language such as \codeql, the state-of-the-art DSL used within the commercial tool LGTM by Semmle/GitHub. At the same time, \fluentql is sufficiently expressive, though, to support the specification of multiple taint-flows which allow the detection of complex security vulnerabilities.

Since Java is still amongst the most widely used languages, we designed \fluentql as an internal Java DSL with a fluent-API design.\footnote{Fluent Interfaces: \url{https://www.martinfowler.com/bliki/FluentInterface.html}} 
This paper presents the syntax and semantics of \fluentql, which is independent of any concrete (static or dynamic) taint analysis.
Our example implementation instantiates \fluentql with two static taint analyses, one based on Boomerang~\citep{popl19spds} and one based on FlowDroid~\citep{flowdroid}.
We explain how these implementations statically approximate the \fluentql semantics.
The implementation is built on top of MagpieBridge~\citep{magpiebridge} and the Language Server Protocol~\citep{lsp}. In result, it can be used in a multitude of editors and integrated development environments (IDEs), including Vim, Eclipse, VSCode, IntelliJ, SublimeText, Emacs, Thea and Gitpod.

We evaluate the usability of \fluentql through a user study with 26 participants (professional software developers, students, and researchers). We compare \fluentql to the more generic \codeql. The results show that software developers perceive \fluentql as easier to use. \fluentql has an excellent System Usability Score (SUS) \citep{sus} of 80,77 (out of 100), whereas (for taint analysis) \codeql has a score of only 38,56\footnote{Interpreting SUS: 0-50 is bad, 51-67 is poor, 68 is an average usability, 69-80,3 is good, > 80,4 is excellent, and 100 is imaginary perfect.}. 
The Net Promoter Score (NPS) \citep{nps} shows that---for the task of specifying taint-flow queries---participants would recommend to others \fluentql over \codeql.  Moreover, we evaluate the applicability of \fluentql by providing specifications of 11 popular vulnerabilities with catalog of small programs. Additionally, we select two vulnerable Java applications (OWASP WebGoat application\footnote{https://github.com/WebGoat/WebGoat} and PetClinic\footnote{https://github.com/contrast-community/spring-petclinic}) and seven real-world Android applications from TaintBench~\citep{taintbench} known with malicious behavior. For all applications, we specified corresponding \fluentql queries and were able to detect most of the expected taint-flows.

To summarize, this paper makes the following contributions:
\begin{itemize}
	\item \fluentql, a new DSL for specifying taint-flow queries, designed to be well usable for software developers. 
	\item A formal definition of the syntax and semantics of \fluentql, the latter independent of any concrete taint-analysis tool. 
	\item An implementation and an empirical evaluation of the usability of \fluentql in comparison to a state-of-the-art DSL for static code analysis. 
\end{itemize}

Our artifact includes the \fluentql tooling, the catalog of queries for popular taint-style security vulnerabilities and the dataset of our user study. It is available anonymously online at \textit{https://fluenttql.github.io/}

We next explain relevant concepts on taint analysis and elicit the requirements for a developer-centric DSL. In Section~\ref{sec-3-fluentql} we present \fluentql with its syntax and semantics, and explain how our static instantiations statically approximate this semantics. In Section~\ref{sec-4-userstudy} we discuss the user study. We discuss related work in Section~\ref{sec-5-relatedwork} and, finally, we conclude in Section~\ref{sec-6-conclusion}.


\section{Requirements for a Taint Analysis DSL}\label{sec-2-motivation}
We next explain the concept of taint analysis with an example and define requirements for a developer-centric DSL for taint analysis.

\listref{list:xss} shows an excerpt of Java code of an HTTP handler. 
The method \emph{doGet} is called upon a GET-request from a web browser when a user changes the password by providing the username, the old password, and the new password. 
The method calls a helper method \emph{changePassword} shown in Listing~\ref{list:nosqli} which verifies the user and changes the database. 
The code in \emph{doGet} contains a potential cross-site scripting vulnerability (XSS)~\citep{cwe79}. 
The username value from the request in the variable \emph{uName} is added to the created HTML page for the response object to inform the user if the password was changed successfully (line~\ref{lstlisting-1-5}). 
There is no sanitization check if the value contains any malicious behavior before it is added to the generated HTML page. 

\begin{lstlisting}[caption={Java code with potential XSS vulnerability (from line~1 to line~5)}, label=list:xss, escapechar=?]{Name}
protected void doGet(%%@RequestParam("user")%%String uName, HttpServletRequest request, HttpServletResponse response) { ?\label{lstlisting-1-1}?
	String oldPass = request.getParameter('oldKey'); ?\label{lstlisting-1-2}?
	String newPass = request.getParameter('newKey');?\label{lstlisting-1-3}?
	if (changePassword(uName, oldPass, newPass))?\label{lstlisting-1-4}?
	response.getWriter().append('<html>... Password changed for user' + uName + '...</html>')); ?\label{lstlisting-1-5}?
	else ?\label{lstlisting-1-6}?
	response.getWriter().append('<html>... 
	Wrong credentials....</html>');?\label{lstlisting-1-7}?
}
\end{lstlisting}

The code in the helper method \emph{changePassword} contains a potential NoSQL injection vulnerability (NoSQLi)~\citep{cwe943}. A single atomic action performs the user authentication and a change of a password in line~\ref{lstlisting-2-11}\xspace in which two database documents (\emph{filter} and \emph{set}), one with \emph{\$where} clause and one with \emph{\$set} clause are executed. To report the taint-flow precisely, both values should be marked as tainted. 
We explain both XSS and NoSQLi vulnerabilities throughout this section.

\begin{lstlisting}[caption={Potential NoSQLi vulnerability (lines~1-3 to line~20)}, label=list:nosqli, escapechar=?]{Name}
protected boolean changePassword(String uName, String oldPass, String newPass) { ?\label{lstlisting-2-1}?
	MongoClient myMongoClient = new MongoClient("localhost", 8990); ?\label{lstlisting-2-2}?
	MongoDatabase credDB = myMongoClient.getDatabase('CREDDB'); ?\label{lstlisting-2-3}?
	MongoCollection<Document> credCollection = credDB.getCollection('CRED', Document.class); ?\label{lstlisting-2-4}?
	BasicDBObject filter = new BasicDBObject(); ?\label{lstlisting-2-5}?
	filter.put('$where', '(username == \"' + uName + '\") \& (password == \"' + oldPass + '\")');  ?\label{lstlisting-2-6}?
	BasicDBObject newPassDoc = new BasicDBObject(); ?\label{lstlisting-2-7}?
	newPassDoc.put('password', newPass); ?\label{lstlisting-2-8}?
	BasicDBObject set = new BasicDBObject();  ?\label{lstlisting-2-9}?
	set.put('$set', newPassDoc);  ?\label{lstlisting-2-10}?
	UpdateResult res = credCollection.updateOne(filter, set); ?\label{lstlisting-2-11}?
	return (res.getMatchedCount() == 1); ?\label{lstlisting-2-12}?
}
\end{lstlisting}

\begin{figure*}[h]
\raggedright
\centering
\includegraphics[width=\textwidth]{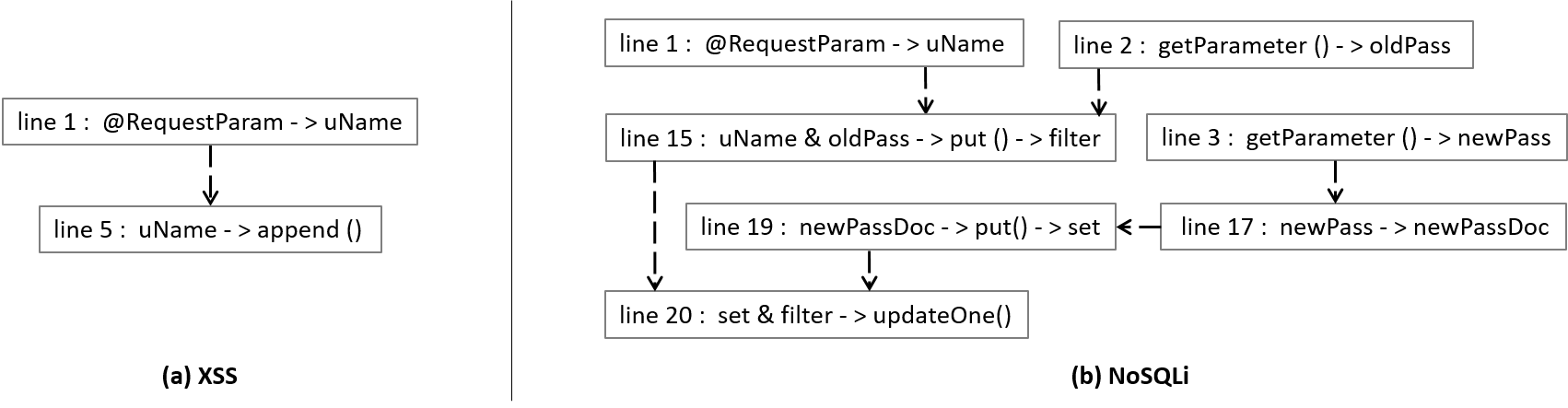}
\captionof{figure}{Data-flow graphs for (a) XSS and (b) NoSQLi vulnerabilities from Listing~\ref{list:xss} and Listing~\ref{list:nosqli}}
\label{fig:example}
\end{figure*}

\subsection{Selection of Sensitive Methods}

To detect such vulnerabilities using a taint analysis, one must configure the analysis with any security-relevant methods (\srm), such as \emph{sources}, \emph{sinks} and \emph{sanitizers}. 

Consider the example of the XSS vulnerability in Listing~\ref{fig:example}. Here, untrusted data flows from the parameter \emph{uName} of the method \emph{doGet} to the sink in line~\ref{lstlisting-1-5}\xspace where method \emph{append()} is called with a string value of a request. 
Figure~\ref{fig:example} (a) shows the data-flow graph extracted from the code. 
To fix this vulnerability, a software developer should apply a \emph{sanitizer} such as \emph{encodeHTML()} to clear potential malicious inputs from the variable \emph{uName} before appending the contents to the HTML string. 
This leads to our first requirement:

\roundbox{\requirement{1}: \emph{The DSL must allow one to express the following security-relevant methods (\srm): source, sanitizer, and sink.} }

\subsection{Selection of In- and Out-Values}
Apart from the selection of the call sites, the actual values flowing in or out of the methods (return values, parameters, and receiver) must be selected. 
For the source of the XSS vulnerability in Listing~\ref{list:xss}, the developer must select the argument value of the first parameter of the method \emph{doGet}. 
At the call to the sink of the vulnerability, the developer needs to provide the possibility to select a parameter of a called method. 

\roundbox{\requirement{2}: \emph{The DSL must allow one to express the data-flow propagation of each \srm to a granularity of single argument, a return value, and a receiver.}}

\subsection{Composition of Taint-Flows}
The presented XSS vulnerability is detected by what we call a ``single-step taint analysis''. It is relatively easy to detect, even manually. 
But many real-world taint-analysis problems comprise a sequence of multiple events. 
For example, consider the NoSQL injection vulnerability in Listing~\ref{list:nosqli} and its data-flow graph in Figure~\ref{fig:example} (b). 

The NoSQLi vulnerability occurs in line~\ref{lstlisting-2-11} when the method \emph{updateOne} is called under the condition that the Mongo database has a record with the username and the old password that matches the values coming from the request object (\emph{uName} in line~\ref{lstlisting-1-1}\xspace  and \emph{oldPass} in line~\ref{lstlisting-1-3}\xspace). 
The value of \emph{filter} contains the document that checks the existing password for the given username by calling the method \emph{put} in line~\ref{lstlisting-2-6}\xspace with a \emph{\$where}-clause. 
The value of \emph{set} contains the document that sets the new password by calling the method \emph{put} in line~18 with a \emph{\$set}-clause. 
When the method \emph{put} is called in line~\ref{lstlisting-2-6}\xspace and line~\ref{lstlisting-2-8}\xspace, the \emph{uName} and \emph{oldPass} taint the \emph{filter} whereas the \emph{newPass} taints the \emph{set}. 
For the taint-flow to be complete, \emph{both} calls to the method \emph{put} must occur before the \emph{set} and \emph{filter} flow to the sink \emph{updateOne()} in line~\ref{lstlisting-2-11}\xspace. 
Thus, we desired a feature to compose complex queries consisting of multiple single-step taint analyses. 

\roundbox{\requirement{3}: \emph{The DSL must allow one to express complex multi-step taint-flow queries.}}

\subsection{Detailed Error Message}
When findings are reported, the analysis tool usually provides a description to the user to help understanding the vulnerability. 
The study of Christakis et al.~\citep{whatdeveloperswant} showed that software developers have difficulties in understanding those descriptions. 
For different vulnerabilities and types of data-flow the DSL shall present the results of the taint analysis with fine-grained error messages that help developers to quickly identify and fix the vulnerability. 
The user that specifies the taint-flow should be able to define a custom error message that can be reported at different locations. 

\roundbox{\requirement{4}: \emph{The DSL must allow one to specify error messages for each type of finding.}}

\subsection{Integration into Developer's Workflow}

Empirical studies show that software developers need static analysis tools integrated in their workflow~\citep{whatdeveloperswant, whydontdevelopers}. 
Most software developers use integrated development environments (IDEs) and prefer static analyses to be directly integrated in the IDE. 
The results of the analysis should be shown within the IDE, preferably visible near the editor for the code. 
Therefore, a DSL designed for software developers should be integrated in this workflow with appropriate tooling and usability.

\roundbox{\requirement{5}: \emph{The DSL must integrate well with the software developers' workflow.}}

\subsection{Independence of Concrete Taint Analysis}

Software developers desire reusing taint-flow specifications for both static and dynamic taint analyses. Moreover, some analysis tools are only part of the continuous integration whereas others can be integrated in different workflows, e.g.\ the IDE. 
To enable reusability of the specifications among different tools, the DSL semantics must therefore be independent of any concrete static or dynamic analysis. Thus, any limitations due to the approximations of the underlying solver are transferred to the results reported by \fluentql. 

\roundbox{\requirement{6}: \emph{The specified taint-flow queries can be reused among existing taint analysis tools, i.e., the DSL is independent of the underlying taint analysis.}}

The NoSQLi vulnerability from the example in this section, can not be detected by default with the existing tools due to its specific structure. Such complex taint-flows require the user to specify a custom query. \fluentql introduced in the next section aims at providing usable and easy approach for mainly software developers specifying custom queries. The existing DSLs are design for experts who have understanding in data-flow analysis, which most developers do not have. Moreover, based on our evaluation of the existing DSLs in Section~\ref{sec-5-relatedwork}, indicates that none of them completely fulfills all requirements.


\section{\fluentql}\label{sec-3-fluentql}

We next define the domain-specific language \fluentql through its abstract  and concrete syntax~\citep{mdsd}. We also define the runtime semantics of \fluentql as independent of a concrete taint analysis. Dynamic taint analyses could faithfully implement the semantics, whereas static taint analyses would seek to soundly approximate it.
Finally, we discuss relevant implementation details. 

\subsection{Concrete Syntax}\label{ssec-3-concrete}

As a concrete syntax for \fluentql, we decided to use a Java fluent-interface syntax. 
Since Java is one of the most popular programming languages, this allows software developers to learn the DSL with little effort. 
Moreover, in interviews with nine software developers~\citep{secucheck}, the authors asked what concrete syntax they would prefer if given the choice of (1) a fluent interface, (2) a graphical syntax, or (3) a textual syntax for taint-flow queries,  
six participants chose the fluent interface, and only two chose the graphical and one the textual syntax. 

In the following, we explain the concrete syntax by specifying the \fluentql queries for the detection of the XSS and NoSQLi code in Listing~\ref{list:xss} and Listing~\ref{list:nosqli}. The specification is presented in Listing~\ref{list:fluentexample1}, where lines \linenumber{fluentql_1}--\linenumber{fluentql_9} contain the \srm declaration and lines \linenumber{fluentql_10}--\linenumber{fluentql_14} contain the taint-flow queries.


\begin{lstlisting}[caption={\fluentql specification for XSS and NoSQLi in Listings~\ref{list:xss} and~\ref{list:nosqli}. To simplify, fully qualified method names are omitted.}, label=list:fluentexample1, escapechar=?]{Name}
Method source1 = new Method("String getParameter(String)").out().return();	?\label{fluentql_1}?
Method source2 = new Method("void doGet(String, HttpServletRequest, HttpServletResponse)").out().param(0); ?\label{fluentql_2}?
MethodSet sources = new MethodSet().add(source1).add(source2); ?\label{fluentql_3}?
Method sanitizer = new Method("String encodeHTML (String)").in().param(0).out().return();?\label{fluentql_4}?
Method reqPropagator1 = new Method("BasicDBObject put(String, String)").in().param(1).out().thisObject(); ?\label{fluentql_5}?
Method reqPropagator2 = new Method("DBObject put(String, DBObject)").in().param(1).out().thisObject(); ?\label{fluentql_6}?
MethodSet reqPropagatorsPut = new MethodSet(). add(reqPropagator1).add(reqPropagator2);?\label{fluentql_7}?
Method sinkXss = new Method("PrintWriter append(CharSequence)").in().param(0); ?\label{fluentql_8}?
Method sinkNoSql = new Method("FindIterable updateOne(BasicDBObject, BasicDBObject)").in().param(0).param(1); ?\label{fluentql_9}?
TaintFlowQuery xss = new TaintFlowQuery().from(source1).notThrough(sanitizer) .to(sinkXss).report("Reflective XSS vulnerability.").at(Location.SOURCE); ?\label{fluentql_10}?	
TaintFlowQuery noSQLi1 = new TaintFlowQuery().from(source1).through( reqPropagatorsPut).to(sinkNoSql).report( "No-SQL-Injection.").at(Location.SINK);?\label{fluentql_12}?	
TaintFlowQuery noSQLi2 = new TaintFlowQuery(); ?\label{fluentql_13}?
noSQLi2.from(source1).through(reqPropagator1).to( sinkNoSql).and().from(source2).through( 
reqPropagator1).to(sinkNoSql).and().from(source1). 
through(reqPropagator2).to(sinkNoSql).report(
"No-SQL-Injection vulnerability with multiple taint-flows").at(Location.SOURCEANDSINK);?\label{fluentql_14}?
}
\end{lstlisting}

In the code there are two potential sources. 
One source is the return value of the \emph{getParameter()} method which in \listref{list:fluentexample1} is specified in line~\linenumber{fluentql_1}. The first argument to the constructor of \syntax{Method()} takes a method signature as a String argument. 
Next, using the fluent interface of \fluentql, we append \syntax{out()} indicating that the method generates a sensitive data-flow. Eventually, by appending \syntax{return()}, we select the return value as the out-value that is generated. 
The other source is the first parameter of the \emph{doGet()} method (line~\linenumber{fluentql_2}) indicated by \syntax{out()} and \syntax{param(0)}. 

The fluent interface of \fluentql allows calling \syntax{out()} or \syntax{in()} on a Method object. 
After \syntax{out()} there has to be at least one more call to \syntax{return()}, \syntax{thisObject()} and/or one or more calls to \syntax{param(int)} with the integer referring to the parameter index of the out-value.
After \syntax{in()} there must be a call to \syntax{thisObject()} and/or one or more calls to \syntax{param(int)}. 

Both sources in line~\linenumber{fluentql_1} and line~\linenumber{fluentql_2} are potential sources for SQLi and XSS, i.e., they are not specific to the vulnerability type. Thus, they are grouped into a \emph{MethodSet} object (line~\linenumber{fluentql_3}). Afterwards, the method \emph{encodeHTML} is specified as sanitizer which is relevant to the XSS vulnerability only.
The method \emph{put()} is a propagator (i.e. only propagates the taint) but a required one, because it has to be called between the source and the sink for this specific vulnerability. 
It can be called with two different parameter types. 
Hence, it is specified twice (lines~\linenumber{fluentql_5} and ~\linenumber{fluentql_6}). They are grouped in the method set \emph{reqPropagatorsPut}. Finally, the sinks are specified (lines~\linenumber{fluentql_8} and \linenumber{fluentql_9}). They are specific to each vulnerability type. 

The taint-flow query for XSS is specified in line \linenumber{fluentql_10} where the class \emph{TaintFlowQuery} is instantiated after which \syntax{from(...)}, \syntax{to(...)}, and \syntax{report(...)} are called. 
For the XSS taint-flow query, the sanitizer is also specified by calling the method \syntax{notThrough(...)}. 
Each of these methods expects an object of type \emph{Method} or \emph{MethodSet}. 

At the end there is a call to \syntax{at(Location.SOURCE)} which is optional and expresses where in the code the report message should be shown. 
\emph{Location} is an enumeration with values \emph{SOURCE}, \emph{SINK}, and \emph{SOURCEANDSINK}.
The taint-flow query can be read as follows: \emph{If there is a taint-flow from the source \syntax{source1} not propagating through the \syntax{sanitizer} and reaching any of the \syntax{sinkXss}, then report a finding with "`Reflective XSS vulnerability"' at the source location.}

For the NoSQLi vulnerability there are two taint-flow queries in \listref{list:fluentexample1}, in lines \linenumber{fluentql_12}--\linenumber{fluentql_14}. 
The object \emph{noSQLi1} will report a finding with a message "`No-SQL-injection vulnerability"' for the source \emph{getParameter}, defined with \emph{source1}, propagating through any required propagator from the set \emph{reqPropagatorsPut} reaching the \emph{sinkNoSql}. 
If applied to the code example from Listing~\ref{list:xss} and Listing~\ref{list:nosqli}, there will be two traces found which will be reported as separate findings. 
The taint-flow from the first parameter of \emph{doGet} carrying the username will be missed. To detect this taint-flow as well, one can use the method set \emph{sources} instead of the single method \emph{source1}.
On the other hand, a taint analysis specified as defined though \emph{noSQLi2} will report a single finding only:
For this specification, the three single taint-flows are joined by a call to \syntax{and()}, which means all separate taint-flows need to occur individually. 

\subsection{Abstract Syntax}\label{ssec-3-abstract}

We discuss the abstract syntax through the meta-model shown in Figure~\ref{fig:metamodel}. 
The DSL has a root node (class \emph{RootNode}) containing all objects. An object of this class represents single instance of the DSL that can contain multiple top level elements. 
The abstract class \emph{TopLevelElement} is a superclass of the main concepts in \fluentql, i.e., the class \emph{Method} and the class \emph{TaintFlowQuery}. 

\subsubsection{Methods} The class \emph{Method} represents a reference to a method from the analyzed code. It contains information about the method signature and the data-flow propagation when that method is called in a given context (conforming to \requirement{1} and \requirement{2}). 
This is expressed through the references to \emph{InputDeclaration} and \emph{OutputDeclaration}. 
A \emph{Method} object has to have one or both \emph{InputDeclaration} or \emph{OutputDeclaration} references. 
An \emph{InputDeclaration} contains an in-value (abstract class \emph{Input}), whereas \emph{OutputDeclaration} contains an out-value (abstract class \emph{Output}). 
In-values can be a parameter of a method call (class \emph{Parameter}) or a receiver of the method (class \emph{ThisObject}). Out-values can be a parameter, a receiver, or a return value (class \emph{Return}). 
In-values flow into the method call and out-values flow out of the method call. 

The class \emph{Method} in combination with the classes \emph{InputDeclaration} and \emph{OutputDeclaration} can model sources, sinks, and sanitizers (\requirement{1}). 
Source is a combination of \emph{Method} and \emph{OutputDeclaration} specifying which values become tainted through a method call. 
Sink is an instance of \emph{Method} and \emph{InputDeclaration} specifying which values must be tainted for the sink to be considered ``reached''. 
Sanitizer is a combination of a \emph{Method} and \emph{InputDeclaration}, specifying which tainted value flowing in the method call will get untainted.

\subsubsection{Required propagators} Required propagators are method calls that have to be on the path between a source and a sink in order for a given vulnerability to be present. For instance the method \emph{put()} in the running example from Section~\ref{sec-2-motivation} has to be on the taint-flow trace from the source to the sink. It only propagates the taint from the in-value to the out-value. 
In \fluentql, a required propagator is modeled as a combination of \emph{Method}, \emph{InputDeclaration}, and \emph{OutputDeclaration}. This model allows propagating out-values once an in-value reaches a method. The analyses that are aware of these methods know how to propagate the data-flow without analyzing them, for example for improving scalability or handling calls for which the source code is not available. 

\begin{figure*}[th]
	\includegraphics[width=\textwidth]{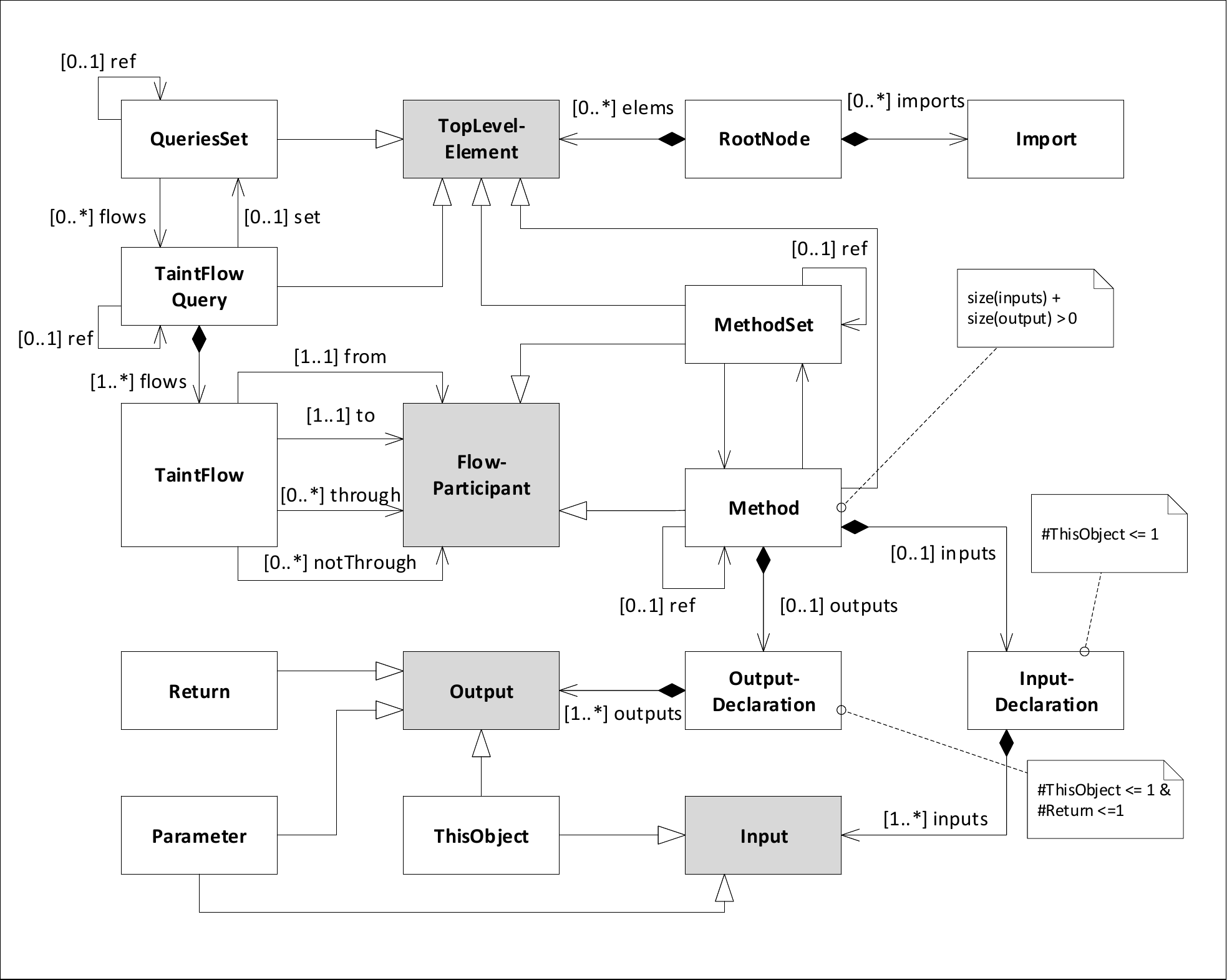}\centering
	\caption{\fluentql meta-model (UML class diagram, gray-filled classes are abstract). The constraints of the cardinalities of the classes are shown as messages, since the semantics of UML class diagram can not express all of them.}
	\label{fig:metamodel}
\end{figure*}

\subsubsection{Taint-flow queries} The class \emph{TaintFlowQuery} represents a taint-flow query. 
It contains all the information one needs to trigger a taint analysis. 
It contains one or more \emph{TaintFlow} objects and a user defined message (\requirement{4}). The class \emph{TaintFlow} has four references to the class \emph{FlowParticipant}. 
The \emph{from} reference defines the set of sources, the \emph{through} reference defines the required propagators, the \emph{notThrough} reference defines the sanitizers, and the \emph{to} reference defines the sinks. 
For any valid \emph{TaintFlow} there should be at least one source and one sink. 
A \emph{FlowParticipant} is either a \emph{Method} or a \emph{MethodSet}, i.e., a collection of methods. 
Similarly, the \emph{QueriesSet} is a collection of taint-flow queries. 

\subsubsection{Imports and reuse} The root node can contain imports from other models defined in other locations. This is modeled via the class \emph{Import}. This allows references of methods and taint-flow queries from different files. The classes \emph{Import}, \emph{MethodSet}, and \emph{QueriesSet} are provided for maintenance, reusability, and structure of \fluentql specifications, enabling software developers to define categories of methods and taint-flow queries and share them (\requirement{5}). As Java internal DSL, the users of \fluentql get all advantages of Java compared to any external DSL or XML/JSON-based DSL, often used in the existing tools. From Java, users can reuse existing abstractions such as packaging, modules, and object-oriented design to improve the maintenance, the readability, and the accessibility of the rules. 

\subsection{Semantics}\label{ssec-3-semantics}

A taint-flow query, an instance of the class \emph{TaintFlowQuery}, is a \fluentql specification that describes which traces of the program should be returned as findings to the user when a given taint analysis is triggered with that taint-flow query.
In the following we define the relevant terms and how \fluentql refers to them. 

We denote \textit{M} to be the set of all method signatures where a signature includes the fully qualified method name, parameter types, and a return type. A sensitive value is a type definition with information about the direction of propagation (in- or out-), and location (return, receiver, or parameter index). Hence, in- and out-values are sensitive values with in- and out-propagation, respectively. 

\begin{definition}\label{def:sm}
	A sensitive method is a tuple \textit{(m,SV)}, where \textit{m}~$\in$~\textit{M} and \textit{SV} is a set of sensitive values. \textit{SV} contains subset $SV_{in}$ for in-values and subset $SV_{out}$ for out-values.
\end{definition}

\begin{definition}\label{def:ta}
	A taint analysis specification \TAS consists of the tuple (\Sources, \Sanitizers, \RPropagators, \Sinks), where 
	\begin{enumerate}
		\item \Sources is a set of sensitive methods $(m,SV_{out})$ for which $SV$ contains at least one out-value, ($SV_{out}\neq\emptyset$),  
		\item \Sanitizers and \Sinks are sets of sensitive methods $(m,SV_{in})$ for which $SV$ contains at least one in-value, ($SV_{in}\neq\emptyset$), and
		\item \RPropagators is a set of sensitive methods $(m,SV_{in},SV_{out})$ containing at least one in-value and one out-value ($SV_{in}\neq\emptyset$, $SV_{out}\neq\emptyset$).
	\end{enumerate}
\end{definition}

Given a taint analysis specification \TAS, some black-box taint analysis $T$ and a program $P$, we assume the execution of $T$ returns a set of traces for the data-flow, i.e., $T(P,\TAS) = \{t_1 \dots, t_n\}$ where each $t_i$ is a data-flow trace. A trace is a sequence of program statements, i.e., $t_i = s_i^1  s_i^2 \dots s_i^n$. For each individual trace $t_i$ it holds that 
\begin{itemize}
	\item the first statement is a source statement, $s_i^1 \in$~\Sources,
	\item the last statement is a sink, $s_i^n \in$~\Sinks
	\item none of the statement $s_i^j$ is a sanitizer,  $s_i^j\notin$~\Sanitizers, and
	\item if \RPropagators is non empty, there exists exactly one element from \RPropagators that appears at statement $s_i^j$, where $j\in\{1,\ldots,n\}$. 
\end{itemize}
Note that in the case the analysis $T$ is a dynamic taint analysis, the set of traces is a singleton set while static analyses, which simulate all possible executions, may generate multiple traces.

\textit{Example:} A \TAS can detect rudimentary data-flows modeled with the class \textit{TaintFlow} from Figure~\ref{fig:metamodel} such as the XSS vulnerability in Listing~\ref{list:xss}. The \textit{TaintFlowQuery} \textit{xss} in line~\linenumber{fluentql_10} specifies a  \textit{TaintFlow} with 
\begin{itemize}[itemindent=7em]
	\item[sources - ] \{\textit{(getParameter(String),}$\outvalue{return}$\textit{), (doGet(String, ...),} $\outvalue{0})$\}
	\item[sanitizers - ] \{\textit{(encodeHTML(String),}$\invalue{0}$\textit{)}\}
	\item[r. propagators - ] \{\}
	\item[sinks - ] \{\textit{(append(CharSequence),}$\invalue{0}$\textit{)}\}
\end{itemize}

\fluentql allows one to specify these sets with respective  syntax elements \syntax{from(...)}, \syntax{notThrough(...)}, and \syntax{to(...)}.
Running a taint analysis with the \fluentql specification for \textit{xss} on the code in Listing~\ref{list:xss} returns the single trace consisting of the two statements\footnote{
	We use line numbers from \listref{list:xss} and \listref{list:nosqli} to represent traces.}  $t_1 = \linenumber{lstlisting-1-1}~\linenumber{lstlisting-1-5}$. 

Additionally, the syntax element \syntax{through(...)} allows to specify the set of \RPropagators. 

For instance, the taint-flow query \textit{noSQLi1} in line~\ref{fluentql_12} specifies a non-empty set \RPropagators. 

\begin{itemize}[itemindent=7em]
	\item[sources - ] \{\textit{(getParameter(String),}$\outvalue{return})$\}
	\item[sanitizers - ] \{\}
	\item[r. propagators - ] \{\textit{(put(String, String),}$\invalue{1}, \outvalue{return}$\textit{), (put(String, BasicDBObject),} $\invalue{1}, \outvalue{return})$\}
	\item[sinks - ] \{\textit{(updateOne(BasicDBObject, BasicDBObject),}$\invalue{0},\invalue{1}),\outvalue{return}$\}
\end{itemize}

The result of the taint-flow query \textit{noSQLi1} is
\[
\textit{Trace$_{noSQLi1}$} = \linenumber{lstlisting-1-2}~ \linenumber{lstlisting-1-4}~ \linenumber{lstlisting-2-1}~  \linenumber{lstlisting-2-6}~  \linenumber{lstlisting-2-11},  ~
\linenumber{lstlisting-1-3}~ \linenumber{lstlisting-1-4}~ \linenumber{lstlisting-2-1}~  \linenumber{lstlisting-2-8}~  \linenumber{lstlisting-2-10}~\linenumber{lstlisting-2-11}\]  and consists of two traces. Each of these traces is reported as a separate finding to the user. A ``simple'' finding is a single trace with a single message.
For instance, the findings of \textit{noSQLi1} are \textit{Findings}$_{noSQLi1}$ = \{ F$_{noSQLi1}^1$, F$_{noSQLi1}^2$\}, where

$F_{noSQLi1}^1$ = (\{\linenumber{lstlisting-1-2}~ \linenumber{lstlisting-1-4}~ \linenumber{lstlisting-2-1}~  \linenumber{lstlisting-2-6}~  \linenumber{lstlisting-2-11}\}, \textit{"No-SQL-Injection vulnerability."})

$F_{noSQLi1}^2$ = (\{\linenumber{lstlisting-1-3}~ \linenumber{lstlisting-1-4}~ \linenumber{lstlisting-2-1}~  \linenumber{lstlisting-2-8}~  \linenumber{lstlisting-2-10}~ \linenumber{lstlisting-2-11}\}, \textit{"No-SQL-Injection vulnerability."})

\noindent
Yet, for more complex queries one can use the \textit{and()} operator, which combines findings over individual traces to a single finding over multiple traces.

\textit{Combining taint-flow queries:} The  \textit{and()} operator allows one to merge multiple \TAS as a single query. This is through an object of type \textit{TaintFlowQuery} (from Figure~\ref{fig:metamodel}) that contains multiple objects of type \textit{TaintFlow}. Formally, the operator computes the cross product of the traces of the individual \TAS.
For example, the taint-flow query \textit{noSQLi2} in line~\ref{fluentql_13}\xspace defines three \TAS specifications:

\begin{itemize}[itemindent=7em]
	\item[sources - ] \{\textit{(getParameter(String),}$\outvalue{return})$\}
	\item[sanitizers - ] \{\}
	\item[r. propagators - ] \{\textit{(put(String, String),}$\invalue{1}), \outvalue{return}$\}
	\item[sinks - ] \{\textit{(updateOne(BasicDBObject, BasicDBObject),}$\invalue{0},\invalue{1})$\}
\end{itemize} 

\begin{itemize}[itemindent=7em]
	\item[sources - ] \{\textit{(put(String, String),}$\outvalue{return})$\}
	\item[sanitizers - ] \{\}
	\item[r. propagators - ] \{\}
	\item[sinks - ] \{\textit{(updateOne(BasicDBObject, BasicDBObject),}$\invalue{0},\invalue{1})$\}
\end{itemize} 

\begin{itemize}[itemindent=7em]
	\item[sources - ] \{\textit{(getParameter(String),}$\outvalue{return})$\}
	\item[sanitizers - ] \{\}
	\item[r. propagators - ] \{\textit{(put(String, BasicDBObject),}$\invalue{1}, \outvalue{return}))$\}
	\item[sinks - ] \{\textit{(updateOne(BasicDBObject, BasicDBObject),}$\invalue{0},\invalue{1})$\}] 
\end{itemize}

The first one returns the trace ``$\linenumber{lstlisting-1-2}~ \linenumber{lstlisting-1-4}~ \linenumber{lstlisting-2-1}~  \linenumber{lstlisting-2-6}~  \linenumber{lstlisting-2-11}$'', the second one returns the trace ``$\linenumber{lstlisting-1-1}~ \linenumber{lstlisting-1-4}~ \linenumber{lstlisting-2-1}~  \linenumber{lstlisting-2-6}~  \linenumber{lstlisting-2-11}$'', and the last one returns the trace ``$\linenumber{lstlisting-1-3}~ \linenumber{lstlisting-1-4}~ \linenumber{lstlisting-2-1}~  \linenumber{lstlisting-2-8}~  \linenumber{lstlisting-2-10}~\linenumber{lstlisting-2-11}$''. Yet, the result of the query \textit{noSQLi2} will be a single finding, \textit{Findings}$_{noSQLi2}$ = \{$F_{noSQLi2}^1$\}, where

$F_{noSQLi2}^1 = (\{\linenumber{lstlisting-1-2}~ \linenumber{lstlisting-1-4}~ \linenumber{lstlisting-2-1}~  \linenumber{lstlisting-2-6}~  \linenumber{lstlisting-2-11},~ \linenumber{lstlisting-1-1}~ \linenumber{lstlisting-1-4}~ \linenumber{lstlisting-2-1}~  \linenumber{lstlisting-2-6}~  \linenumber{lstlisting-2-11},~ \linenumber{lstlisting-1-3}~ \linenumber{lstlisting-1-4}~ \linenumber{lstlisting-2-1}~  \linenumber{lstlisting-2-8}~  \linenumber{lstlisting-2-10}~\linenumber{lstlisting-2-11}\}$, \textit{"No-SQL-Injection vulnerability with multiple taint-flows."}$)$.

\textit{Calculating traces:} By its definition, \fluentql has a precise runtime semantics. However, when applied in static context, the traces need to be approximated by the underlying data-flow engine. Thus, reported traces of different tool implementations can differ. 

To explain the precise runtime semantics for traces construction, we define a taint analysis core language, in similar fashion to previous works~\citep{dynamistaintsem1, dynamictaintsem2}. 
Though simple, the core language covers relevant statements that can be mapped one-to-one with Java statements. The statements are listed in Table~\ref{table:corelanguage}. 
A program of the language contains a sequence of statements with line number. For simplicity, we decided to exclude method calls from the language. These can be compiled to the language by storing the memory address of the return statement and transferring the control flow. 
This rule is not applied to the four statements in Table~\ref{table:corelanguage} which are special method calls. 

\begin{table}[ht]
	\small
	\centering
	\caption{Statements of the core language for constructing \fluentql traces}
	\setlength\tabcolsep{.5mm}
	\begin{tabular}{c l}
		\textbf{ Statement } & \textbf{Description}  \\
		\hline
		src(x) & call to a sensitive method \textit{(m,$SV_{out}$)} $\in$\Sources with sensitive parameter $x$\\ 
		snk(x) & call to a sensitive method  \textit{(m,$SV_{in}$)} $\in$\Sinks with leaked parameter $x$\\ 
		san(x) & call to a sensitive method \textit{(m,$SV_{in}$)} $\in$\Sanitizers sanitizing parameter $x$ \\ 
		rpr(x) & call to a sensitive method  \textit{(m,$SV_{in}$,$SV_{out}$)} $\in$\RPropagators  \\
		$x = y$ & assignment  \\
		$x = y.f$ & field load  \\
		$x.f = y$ & field store  \\
		$x = a[i]$  & read from array at index i  \\
		$a[i] = x$ & write to array at index i  \\ 
		$skip$ & skip and continue  \\ 
	\end{tabular}
	\label{table:corelanguage}
\end{table}

We denote variables with $x$ and $y$, a field of an object with $f$, and an i-th index of an array with $a[i]$. 
We model all memory locations through a shadow heap: The shadow-heap values for a memory location $v$ is \textit{true} if the value is tainted and \textit{false} otherwise.
The execution context $\Sigma$ has the parameters listed in Table~\ref{table:contextparams}. 
$\Sigma.\Delta[x]$ stores the current taint value of variable $x$. We write $\Sigma \vdash x \Yields v$ to extract that value into $v$.
Similarly, notations like $src(x) \vdash (m,SV)$ extract the method $m$ with its sensitive values $SV$, when a method call $src$ is matched. 
Additionally, $\Sigma$ stores all traces $t \in T$ that will be created during the execution. $t$ is a sequence of statements (which we here denote by line numbers).

\begin{table}[htp]
	\small
	\centering
	\caption{Statements of the core language for constructing \fluentql traces}
	\setlength\tabcolsep{.5mm}
	\begin{tabular}{c p{10cm}}
		\textbf{Parameter} & \textbf{Description}  \\
		\hline
		$\Delta$ & match a variable, a field, an array element or a sensitive value to its taint value \\ 
		$\lambda$ & match a given statement to its line number \\
		$\theta$ & returns the set of all traces created \\
	\end{tabular}
	\label{table:contextparams}
\end{table}

\begin{figure*}[th]
	\includegraphics[width=\textwidth]{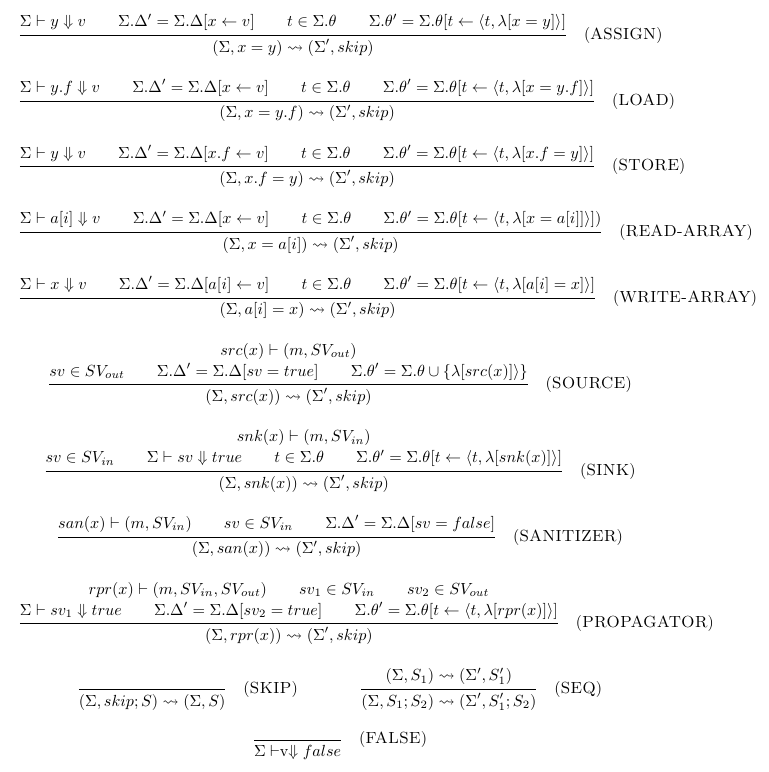}\centering
	\caption{Inference rules of the operational semantics of the traces construction in \fluentql}
	\label{fig:semantics}
\end{figure*}

\autoref{fig:semantics} shows \fluentql's semantics through inference rules. We use a syntax akin to the one used by Schwartz et al.~\citep{dynamistaintsem1}. The semantics essentially define a regular dynamic taint analysis which, as side-effect collects un-sanitized traces from sources to sinks.
For instance, given the statement $x = y$, the ASSIGN rule's computation comprises four parts. First, $\Sigma \vdash y \Yields v$ evaluates and extracts the taint value $v$ for variable $y$. Due to the assignment, the rule updates the taint value of $x$ with $v$. The rule then also extracts each trace in $t \in \theta$ and adds to it the current statement, identified by its line number. 
The rules for load/store and array accesses are equivalent. 
The rule SOURCE creates a new trace and taints the out-value, the rule SINK gracefully terminates a trace by untainting the sensitive value. The rule SANITIZER also discontinues the tracing. The rule PROPAGATOR taints the out-value if the in-value is tainted. SKIP advances to the next statement whereas SEQ enables the progression of the semantics covering the recursive case. 
The semantics must additionally enforce one aspect that we found hard to capture with inference rules: for such \fluentql specifications that define required propagators, the taint analysis must ensure to report only such traces that actually contain all required propagators. Finally, the notion of user-defined message is skipped in the formal semantics due to simplicity, but we explain it in the following through our example. 

\textit{Report message:} As seen in the previous examples, the queries specified in \fluentql contain a user-defined message which is added to each finding (\requirement{4}). 
In the concrete syntax, the mandatory syntax element \syntax{report(...)} takes the string message as an argument. 
Optionally, the user may specify the location for the reporting message by using the syntax element \syntax{at(...)}. 
As an argument, the enumeration \textit{Location} can be used, which contains three elements \textit{SOURCE, SINK} and \textit{SOURCEANDSINK}. 
\textit{SOURCE} and \textit{SINK} define that the reporting message should be shown at the source and the sink location respectively. For \textit{SOURCEANDSINK} the message should be shown at both source and sink location in the code. If the finding has multiple traces then the reporting message is shown for each trace individually. 
For example, for \textit{noSQLi2} the error message will be shown at each source and sink location, i.e., lines 1, 2, 3, and 20, because \textit{SOURCEANDSINK} is used in the query specification (Listing~\ref{list:fluentexample1}, line~\ref{fluentql_14}). 
This information can be used by tools for visualization purposes. E.g. an IDE plug-in may display error markers in the editor at the source location, the sink location, or both. 

\textit{Usability versus expressiveness:} \fluentql is a DSL for users without deep expertise in static analysis as most software developers. Its purpose is to enable users specify a custom taint analysis for their codebase and detect many popular security vulnerabilities. Hence, the usability and simplicity of the language is the primary aim. A trade-off to this design decision is the lower expressiveness when compared to some existing DSLs such as \codeql. \fluentql does not provide the users a fine-grained manipulation of the abstract syntax tree (AST). Such expressive DSLs are used by program analysis experts. This fine-grained AST manipulation, can be useful for writing more compact code. Nonetheless, as our evaluation in Section~\ref{sec-4-userstudy} shows, most popular security vulnerabilities can be expressed in \fluentql. This is due to the fact that the relevant data being tracked by the analysis is impacted only by specific method calls within the program, which is the case for most security vulnerabilities. Compared to \codeql, in \fluentql, other language constructs than method calls currently can't be modeled. However, as discussed in Section~\ref{ssec-4-rq3}, extending \fluentql with new language constructs is possible without significant semantic changes. On the side of expressiveness, \fluentql has a support of \requirement{3} which is only partially supported by other DSL as can be seen later in Section~\ref{sec-5-relatedwork}. Complex multi-step taint-flow queries are in particular relevant for stored versions of SQLi and XSS vulnerabilities. Finally, \fluentql only support taint analysis, whereas other DSLs like \codeql support additional types of analyses, such as value analysis. 

\subsection{Implementation}\label{ssec-3-impl}

We implemented \fluentql as an internal Java DSL which can be easily used in any Java project by implementing the interface \emph{FluentTQLSpecification}. Hence, any Java editor can be used to write and edit \fluentql queries. 

Additionally, we implemented a server using the MagpieBridge framework~\citep{magpiebridge} that can trigger, execute the analysis, and return the results to the IDE. \fluentql is implemented as a standard Java library using the builder pattern to allow method chaining as user interface. All queries need to be implemented within a class that implements the interface \emph{FluentTQLSpecification}. Using the Java classloader the classes are located and the queries correctly loaded and provided as input to the analysis. 

\begin{figure*}[th]
	\includegraphics[width=0.7\textwidth]{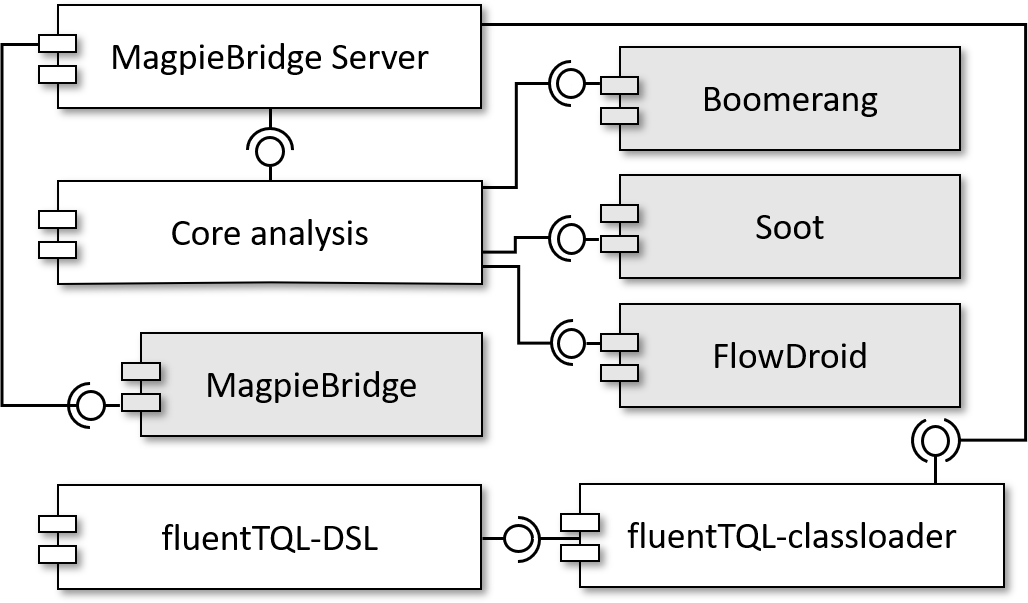}\centering
	\caption{Component diagram of the \fluentql imeplementation as MagpieBridge server (gray components are external, white components are internal)}
	\label{fig:archi}
\end{figure*}

As we rely on MagpieBridge, we support IDEs that support the Language Server Protocol~\citep{lsp} such as Vim, Eclipse, VSCode, IntelliJ, and many more. The MagpieBridge server uses the Language Server Protocol to notify the IDE for available results. Figure~\ref{fig:archi} shows a component diagram of our implementation. The core analysis uses Soot as an underlying static analysis framework responsible for providing the main data structures, such as control-flow graph and call graph. Solvers such as Boomerang and FlowDroid provide interface for starting a taint analysis. The core analysis utilizes this to execute the semantics of the \fluentql queries as described previously in \fluentql's semantics. The core analysis matches the solver's APIs with the \fluentql queries that are executed. For complex queries it breaks them into simple taint-flows which are independently solved by the underlying solver and their results merged afterwards. The queries are loaded through the \fluentql-classloader into the MagpieBridge-Server. 

Our implementation uses the standard IDE features: errors view, editor markups, and notifications to display the results from the analysis directly in the IDE. Additionally, it provides a configuration page where the user can filter the queries and the entry points used for the call graph used by the analysis.

To instantiate \fluentql with concrete analyses, we first implemented a taint analysis built on top of the Boomerang solver~\citep{popl19spds}, an efficient and precise context-, flow-, and field-sensitive data-flow engine with demand-driven pointer analysis. Boomerang provides an API to query all traces from given seeds. The API of the seed is expressible to cover the \fluentql semantics of the sensitive methods. However, the basic API of Boomerang does not support sanitizers, nor required propagators. To support the sanitizers we transformed the bodies of the sanitizers to empty, which is a terminal case of the Boomerang data propagation solver. To support required propagators, we break the \TAS specification to multiple \TAS specifications containing only sources and sinks. A \TAS with required propagator is broken to two \TAS where the first one has the original source and the required propagator as sink, whereas the second one has the required propagator as source and the original sink as sink. Boomerang returns the traces of the individual \TAS, and our implementation merges them. There is no explicit well-formed check in our implementation. However, we implemented the taint analysis with Boomerang on ourselves and we, therefore, trust its correctness with respect to the semantics of the constructed traces. However, future implementation with other solvers, should also include a well-formed check. 

Moreover, we instantiated \fluentql with the existing taint analysis of FlowDroid~\citep{flowdroid}. This, however, was not possible without limitations. Specifically, the default component for defining sources and sinks in FlowDroid is limited and supports only return as out-value of sources and parameter index as in-value of sinks. This can be extended by adding new implementation of the \textit{SourceSinkManager}, which we left as future work. Sanitizers by default are not supported, but we applied the same solution as in our Boomerang implementation, whereas required propagators are not supported and requires either extension of the taint analysis or post-processing of the findings which we also consider as future work.  

Finally, both instances of \fluentql have some limitations in the way the traces are constructed and reported. Since \fluentql has precise runtime semantics, it is expected that static analysis engines like Boomerang and FlowDroid will approximate. In particular, both engines will unsoundly underapproximate the constructed traces. For example, both apply different strategies for merging conditional paths of the program. Thus, these limitations are part of our implementation, too.


\section{Evaluation}\label{sec-4-userstudy}

We evaluated the usability of \fluentql by conducting a comparative user study between \fluentql and \codeql. We chose \codeql because it is part of LGTM, a state-of-the-art security tool, which has, in our perspective, very good tool support and the query specifications are open-source. There is also an Eclipse plugin, a web console for queries, and integration with GitHub, a popular versioning system among developers. Additionally, we evaluated the applicability of \fluentql by specifying queries for different set of applications: a catalog of eleven Java programs, each demonstrating different security vulnerability, the deliberately insecure application OWASP WebGoat aiming to teach developers about relevant security vulnerabilities, an insecure version of the Spring Demo application PetClinic, and randomly selected five real-world Android apps with known malicious taint-flows part of TaintBench~\citep{taintbench}. All selected applications have known expected taint-flows that can be used to evaluate how does the analysis perform in finding real vulnerabilities. We answer the following research questions:
\begin{itemize}
	\item \RQ{1} How usable is \fluentql for software developers? 
	\item \RQ{2} How does \fluentql compare to \codeql for specifying taint-flow queries for taint-style security vulnerabilities?	
	\item \RQ{3} Are \fluentql syntax elements sufficient to express queries for popular taint-style security vulnerabilities?
	\item \RQ{4} Can \fluentql express and detect the known security vulnerabilities Java/Android applications?
\end{itemize}

To answer the research questions, we use corresponding metrics. For \RQ{1}, we use the System Usability Scale and Net Promoter Score. The same metrics are also used in \RQ{2} to compare both DSLs. Additionally, we measure the time needed for the participants to complete the given tasks. We count only the solutions which are complete queries. The partial solutions are not counted due to the nature of the task. In similar realistic scenario, incomplete queries will not return results from the tools. For \RQ{3}, we evaluate how each \fluentql construct contributes in specifying the most popular Java security vulnerabilities. Moreover, we identify security vulnerabilities for which \fluentql can not express the required constructs. Finally, for \RQ{4}, we count how many of the expected taint-flows in the selected applications are found when \fluentql runs with adequate queries. 

The following subsection explains our methodology for the user study used to answer \RQ{1} and \RQ{2}. The next subsections discuss the results of each research question individually. Finally, we discuss threats to validity. 

\subsection{Methodology}\label{ssec-4-methodology}

\textit{Setup:} The user study was conducted over a set of teleconferences where each participant shared the screen. Each study took on average 80 minutes. The session was recorded for post-processing purposes. We invited 35 software developers to take part in the study, from which 26 accepted the invitation, referred to as P01-P26. We invited professional developers via our contacts from the industry as well as researchers and master level students. Additionally, we asked three students to participate in a test session, which helped us to estimate the time and adjust the difficulty of the tasks. 

Due to the limited number of participants, we chose a within-subjects design. Hence, each participant worked in Eclipse with available tool support for both DSLs. The \fluentql implementation used the more versatile instantiation based on Boomerang. To avoid any bias, we referred to the DSLs by DSL-1 and DSL-2. Initially, the participants received a project with all files needed for the practical part. The moderator gave an introduction to taint analysis and showed a Java code example with an SQL injection vulnerability~\citep{cwe89} to make sure that the participant understands the required concepts such as source, sanitizer, required propagator, and sink. Then, the exercises for DSL-1 and DSL-2 followed. 

Each exercise consisted of a tutorial and a task. The tutorial for each DSL was based on the SQL injection vulnerability. Then, the participants had ten minutes to write a specification in the same DSL for a new vulnerability explained by the moderator. We chose the vulnerability types open redirect~\citep{cwe601} for \fluentql and cross-site scripting~\citep{cwe79} for \codeql. For either type, we selected an example with the same pattern in form of source-sanitizer-sink. This ensures that writing a specification for each vulnerability is equally hard, i.e., the effort is the same regardless of the vulnerability. 

For each vulnerability type, we provided a Java code example as a reference. The participant was allowed to use any of the files provided that included the Java classes and the files with example specifications of \fluentql and \codeql. For each task, we additionally provided a file with a skeleton code in which the participant wrote the solution. During the tasks, the participants were allowed to ask questions for clarification. 

After the tasks, we let the participants fill a web form. The moderator guided the participant in the discussion and collected the data for the questionnaire. 

\textit{Questionnaire:} In total the questionnaire asked 28 questions, of which two are of open type and optional (Q26 and Q27). The complete list of questions is part of our artifact. Each of the questions asks for feedback for each DSL by the participant. From the 26 mandatory questions of closed type, 4 are informational, 20 are related to the System-Usability-Scale (SUS)~\citep{sus}, and two are related to the Net Promoter Score (NPS)~\citep{nps}. The SUS value is a usability metric that can be calculated with ten simple questions in a predefined format. The SUS-related questions (Q4-Q23) are the same ten questions per DSL with answering options on agreement scale from one to five. SUS expresses usability of a single DSL. Hence, for comparison we use the same questions for each DSL. The NPS metric expresses how likely the participant would recommend something to a colleague.  To calculate a value, NPS identifies so-called promoters and detractors among the participants. The NPS-related questions (Q24-Q25), ask for the likelihood of DSL1 being recommended over DSL2 for the task of specifying taint-flow queries and vice versa. The informational questions ask about participant coding experience (Q1), security expertise (Q2), willingness to learn a new DSLs (Q3), and preferred way of learning new languages (Q28).

\textit{Participants:} The study population with 26 participants is larger than the size of related studies that have been performed earlier, e.g. 10 in~\citep{howdevelopers}, 12 in~\citep{whycantjohnny}, and 22 in~\citep{explaining}. We chose participants with a diverse background. Ten of them are professional developers, six are computer science students on the master level, and ten are researchers in computer science. The participants have different experiences in programming. Twelve of the participants have 10+, nine have 6-10, four have 3-5, and one has 1-2 years of programming experience. They rated their experience with security vulnerabilities. Three consider themselves as beginners, 16 have basic knowledge, five regularly inform themselves about the topic, and two consider themselves as experts. 

\textit{Statistical tests:} Along with the reported data and metrics, we perform relevant statistical tests. As a within-subject design our collected data is paired, i.e, for each participant we have one set of collected data. Exception are the SUS and NPS metrics which are aggregated among all participants. The limitation of this design is the possibility of carryover effects, such as learning effects. The main treatment variable is the \textit{technique},  stating which DSL was used to solve each task (nominal data). In addition, we have an independent crossover treatment variable, the choice of DSL for the first task (binomial data). The background variable are: years of coding experience (ordinal data), position (nominal data), and security experience (nominal data). Finally, we have two effort variables, one for the outcome of each task (binomial data) and the time (ratio data). As most of the data is nominal and ordinal, we used only non-parametric statistical tests. Bellow we report individually each selected test and the results. We used the significance level $\alpha$ = 0.05 for all the tests.


\subsection{\textit{\RQ{1} Usability of \fluentql} }\label{ssec-4-rq1}
\fluentql was positively received by the participants of our user study. It received an excellent System Usability Score of 80,77 on a scale from 0 to 100 where 68 is considered to be an average usability and 100 is imaginary perfect. 

Using the null hypothesis "\textit{\fluentql is usable (SUS is bigger then the hypothetical value of 68)}", we select the Wilcoxon test (the data is ratio but without normal distribution). The test accepts the hypothesis with statistical significance and large effect size (>0.5). For the given task, 20 out of 26 participants have finished with a correct solution in 10 minutes (on average 472 seconds, with $\sigma=$99,05). Table~\ref{table:participants} shows the exact time in seconds for each participant. In the open questions (Q26-Q27), many of the participants gave additional feedback what they like and what they would improve in \fluentql. Most of the participants said that they can learn the language very easily, one of them said "`\textit{with simple tutorial, I can learn it (\fluentql) even without an expert. (...) it was very intuitive}"' and other said "`\textit{I didn't have to learn a lot}"'. Few participants mentioned that they like that the queries are compact and have the right level of abstraction. 

We noted a few points that many participants disliked. Most dislike that the method signatures are specified as a string value. One participant said "`\textit{method calls are prone to typos or cumbersome to create}"'. For this, we already added a check in the editor to inform the users if their string is an invalid method signature. We support Java and Soot signatures. We even plan to add suggestions for existing methods from the workspace to the code completion feature of the editor. Some participants gave suggestions for improving the names of some keywords. For example the class \emph{ThisObject}, which in \fluentql is called with \emph{thisObject()}, was earlier called \emph{This} and confused many participants with the \emph{this} keyword in Java. 

\roundbox{\fluentql as a new DSL is found to be very usable. The participants of our user study gave a score of 80,77 on the System Usability Score system.}

\subsection{\textit{\RQ{2} Comparison of \fluentql and \codeql} }\label{ssec-4-rq2}
In terms of usability, with a SUS value of 38,56 \codeql is perceived with bad usability. Using the null hypothesis "\textit{\codeql is not usable} (SUS is smaller than the hypothetical value of 68)",  we select the Wilcoxon test. The test accepts the hypothesis with statistical significance and large effect size (>0.5). On the questions how likely will the participant recommend one DSL over the other for the task they were given (Q24-Q25), \fluentql over \codeql has a Net Promoter Score value of 30,77,  whereas \codeql over \fluentql has a value of -86,96, where on the scale from -100 to 100, positive values are considered good. It follows that for specifying taint-flow queries, participants would more likely recommend \fluentql over \codeql. 

\begin{table}[H]
	\small
	\centering
	\caption{List of participants: coding experience and position, time in solving each task and DSL used in the first task (X means the participant did not solve the task in 10 minutes)}
	\setlength\tabcolsep{.5mm}
	\begin{tabular}{c|c|c|c|c|c|c}
		& \begin{tabular}{@{}c@{}}Coding \\ (years)\end{tabular} & Position& \begin{tabular}{@{}c@{}}Security \\ experience\end{tabular}& \begin{tabular}{@{}c@{}}\fluentql \\ (seconds)\end{tabular}  & \begin{tabular}{@{}c@{}}\codeql \\ (seconds)\end{tabular}  & 1st DSL  \\ \hline
		P01 & 3-5 & developer & basic & 554 & X & \fluentql  \\
		P02 & >10 & developer & basic &499 & X & \fluentql  \\
		P03 & 6-10 & student & expert & 482 &  588 & \fluentql  \\
		P04 & >10 & researcher & basic & 560 & 590 & \codeql \\
		P05 & >10 & researcher & basic & X & 591 & \fluentql \\
		P06 & >10 & researcher & advanced & 544 & 562 & \fluentql \\
		P07 & 3-5 & researcher & basic & X & 595 & \fluentql \\
		P08 & >10 & student & advanced & 449 & 495 & \codeql \\
		P09 & 6-10 & student & basic & X & 587 & \codeql \\
		P10 & >10 & researcher & basic & 545 & 567 & \codeql \\
		P11 & 1-2 & researcher & beginner & 558 & 585 & \codeql \\
		P12 & 6-10 & researcher & basic & X & X & \fluentql \\
		P13 & 3-5 & researcher & beginner & 473 & 541 & \codeql \\
		P14 & 6-10 & researcher & basic & 305 & 434 & \codeql \\
		P15 & 6-10 & researcher & basic & 571 & X & \fluentql \\
		P16 & 6-10 & student & beginner & 412 & 558 & \codeql \\
		P17 & >10 & developer & basic & X & X & \fluentql \\
		P18 & >10 & developer & basic & 328 & 600 & \codeql \\
		P19 & 6-10 & developer & basic & 594 & X & \fluentql \\
		P20 & 6-10 & developer & expert & 375 & 492 & \codeql \\
		P21 & 6-10 & student & basic & 455 & 467 & \codeql \\
		P22 & >10 & developer & advanced & X & X & \codeql \\
		P23 & >10 & developer & advanced & 507 & 600 & \fluentql \\
		P24 & >10 & developer & advanced & 206 & 425 & \codeql \\
		P25 & 3-5 & student & basic & 531 & X & \fluentql \\
		P26 & >10 & developer & basic & 492 & X & \fluentql \\
		\hline
	\end{tabular}
	\label{table:participants}
\end{table}

\begin{lstlisting}[caption={\codeql specification for XSS}, label=list:codeql, escapechar=?]{Name}
	class XSSConfig extends TaintTracking2::Configuration {
		XSSConfig() { this = "XSSConfig" } 
		override predicate isSource(DataFlow::Node source) { source instanceof RemoteFlowSource } ?\label{codeql_1}?
		override predicate isSink(DataFlow::Node sink) { sink instanceof XssSink }?\label{codeql_2}?
		override predicate isSanitizer(DataFlow::Node node) {
			node.getType() instanceof NumericType or node.getType() instanceof BooleanType} ?\label{codeql_3}?
	}
	from DataFlow2::PathNode source, DataFlow2::PathNode sink, XSSConfig conf ?\label{codeql_4}?
	where conf.hasFlowPath(source, sink) ?\label{codeql_5}?
	select sink.getNode(), source, sink, "Cross-site scripting due to $@.", source.getNode(), "user-provided value" ?\label{codeql_6}?
}
\end{lstlisting}

To compare both languages, let us consider the \codeql example for XSS in Listing~\ref{list:codeql}. 
This is a solution for the task given to the participants. 
The query (lines~\ref{codeql_4}-~\ref{codeql_6}) consists of three sections, \emph{from}, \emph{where}, and \emph{select}. 
In the \emph{from} section, the user defines objects from predefined or self-defined classes. 
In the \emph{where} section, constraints are defined that may also contain calls to predicates. 
In the \emph{select} section, the results of the query are defined. 
For taint analysis, \codeql provides a module. The class \emph{XSSConfig} extends from the configuration class for taint analysis where the sources, sanitizers, and sinks are defined. 
Additionally, the classes \emph{RemoteFlowSource} and \emph{XssSink} are provided and can be used to detect sources and sinks for XSS. 
The stub code with relevant imports given to each participant contained information that these classes exist and can be used. 
A user who needs other \srm that the provided classes cannot detect, will need to write a new implementation. 
Note that the provided classes \emph{RemoteFlowSource} and \emph{XssSink} will match more sources and sinks than the \fluentql query solution. To have an equivalent query as the one in \fluentql, the participants would have to write additional code for the  \emph{isSource} (Line~\ref{codeql_1}) and  \emph{isSink} (Line~\ref{codeql_2}) methods instead of using the provided classes. 

Few participants mentioned the amount of code they would need to write in \codeql is large. One participant said, \textit{"'...way too much code to get to the actual thing that needs to be written."'}. 

Furthermore, we observed how each participant performed in solving the tasks. The task with \codeql was solved by 17 participants, compared to 20 with \fluentql. Fourteen participants solved both tasks. However, on this data, the Fisher's test (selected due to binomial small sample) did not indicate a statistical significance. We measured the time each participant needed for each task, which is given in Table~\ref{table:participants}. On average participants solved the task with \codeql in 546 seconds ($\sigma=$57,89), which is by 13,4\% slower than with \fluentql. Using the Wilcoxon test we found a statistical significance for the null hypothesis with a small effect. 

We performed few additional Wilcoxon tests for the impact of the background variables, i.e. \textit{Coding}, \textit{Position}, and \textit{Security experience}. None of these tests showed a statistical significance of the null hypothesis which tested whether the variable impacts the time of solving the tasks. Finally, we look into the outcome of each task. The null hypothesis is "\textit{The order of the tasks impact the output}". We used the two-way ANOVA~\citep{anova} test, which did not show a statistical significance. Hence, we reject the null hypothesis and accept the alternative one stating that the order of the tasks does not impact the outcome. 

\roundbox{While \codeql is more expressive DSL for multiple types of static analyses, \fluentql is more preferred among software developers due to its user-friendliness. \codeql scored a bad SUS value of 38,56. On the NPS system, \fluentql is preferred over \codeql with a score of 30,77, whereas \codeql is preferred over \fluentql with a negative value of 86,96. When specifying a taint-flow for given known vulnerability, the participants in our user study were 13,4\% faster when using \fluentql compared to \codeql.}

\subsection{\textit{\RQ{3} Expressiveness }}\label{ssec-4-rq3}
To evaluate whether \fluentql syntax elements are sufficient to express popular Java taint-style vulnerabilities, we created a catalog with Java code examples accompanied by \fluentql specifications. 
The catalog contains eleven types of security vulnerabilities (Table~\ref{table:catalog}). 
Each Java code example has a variant with and without sanitization. 
The catalog demonstrates different language syntax elements of \fluentql and how they can be used for specifying vulnerabilities. 
The Java examples and the \srm are manually collected from several sources including the Mitre \citep{cwe} and OWASP \citep{top10} databases, OWASP benchmark project\citep{owaspbench}, and other publicly available \srm lists \citep{susi, sus, swan, joanaudit}. 

\begin{table}[H]
	\small
	\centering
	\caption{List of vulnerability types implemented in the \fluentql catalog (so - sources, sa - sanitizers, rp - required propagators, si - sinks)}
	\setlength\tabcolsep{.5mm}
	\begin{tabular}{l|c|c|c|c|c|c}
		Vulnerability type& flows & so & sa & rp & si & \srm \\ \hline
		SQL injection~\citep{cwe89} & 3 & 13 & 3 & 6 & 10 & 32\\
		XPath~\citep{cwe643} & 1 & 12 & 1 & 0 & 12 & 25 \\
		Command injection~\citep{cwe77} & 1 & 12 & 1 & 1 &  1 & 15 \\
		XML injection~\citep{cwe91} & 1 & 12 & 1 & 0 & 4 & 17\\
		LDAP injection~\citep{cwe90} & 1 & 12 & 1 & 0 & 8 & 21 \\
		Cross-site scripting~\citep{cwe79} & 2 & 13 & 1 & 1 & 3 & 18 \\
		Open redirect~\citep{cwe601} & 2 & 13 & 1 & 0 & 2 & 16\\
		NoSQL injection~\citep{cwe943} & 2 & 5 & 2 & 3 & 2 & 12 \\
		Trust boundary violation~\citep{cwe501} & 1 & 12 & 1 & 0 & 1 & 15\\
		Path traversal~\citep{cwe23} & 2 & 12 & 1 & 1 & 2 & 16\\
		Log injection~\citep{cwe117} & 2 & 12 & 1 & 1 & 4 & 18 \\
		\hline
		Total (unique): & 18 & 46 & 14 & 13 & 49 & 122 \\
		\hline
	\end{tabular}
	\label{table:catalog}
\end{table}

Many of the taint-style vulnerabilities from the Mitre and OWASP databases can be modeled with single taint flow queries. Yet, we found some examples such as the \emph{noSQLi2} query in Listing~\ref{list:fluentexample1} where the and() operator is needed. 

Taint flows that require multiple intermediate source-sinks steps were necessary for the specification of many taint flows, i.e., the feature of multi-step taint analysis is ubiquitous. For example, the OWASP Benchmark test 00001\footnote{\url{https://github.com/OWASP/Benchmark/blob/master/src/main/java/org/owasp/benchmark/testcode/BenchmarkTest00001.java}} contains Path Traversal vulnerability \citep{cwe23}. A \textit{File} object is constructed using a \textit{String} parameter as location to the file. If the \textit{String} is user-controllable, i.e., tainted, and the \textit{File} object is passed to a \emph{FileInputStream} constructor, a path traversal vulnerability occurs. 
The file constructor in this case is a required propagator that ensures the order of \srm calls.

When it comes to the \srm specifications, we observed that most of the sources have a \emph{Return} object as an out-value. For sinks, most of the in-values are \emph{Parameter} objects.

Additionally, we inspected the vulnerability types (known as Common Weakness Enumerations - CWEs) in the SANS-25 list \citep{top25}. 17 of 25 vulnerability types can be expressed as taint-style. 13 of those can be modeled in \fluentql. 

The remaining four CWEs are: CWE-119, CWE-787, CWE-476, and CWE-798.
Both, CWE-119 and CWE-787 are related to buffer overflows, which do not apply to Java. The CWE-476 cannot be expressed because the potential sources are \texttt{new}-expressions, which cannot currently be modeled.
Also, constant values cannot currently be modeled as potential sources which is needed for CWE-798 where these values should be detected as hard-coded credentials. 
Extending \fluentql to support \texttt{new}-expressions and constant values is possible in the abstract syntax by modeling them with a new class that extends the class \emph{FlowParticipant}. The semantics needs to be extended to define how these values will be detected and define appropriate concrete syntax. 

Even though our implementation of \fluentql if bound to Java only, \fluentql can express taint-style vulnerabilities in other languages too. To specify a query for other languages, the only requirement is that the sources, sanitizers, and sinks are defined as method calls. Similar to Java, \fluentql can be adapted to work for C/C++, C\#, other JVM-hosted languages and cover a wide range of taint-style vulnerabilities. In languages such as JavaScript, the coverage of vulnerabilities is smaller since the sources and sinks are often not method calls.

\roundbox{\fluentql is able to express all taint-style vulnerabilities in which the key constructs of the taint-flows are method calls. Our implementation shows that at least 11 types of security vulnerabilities can be specified with \fluentql. These are the most popular security vulnerabilities for Java. Theoretically, one can express many more.}

\subsection{\textit{\RQ{4} Analyzing Java/Android Applications} }\label{ssec-4-rq4}

To answer \RQ{4} we ran \fluentql queries on two Java applications, OWASP WebGoat and PetClinic, and seven Android applications from TaintBench~\citep{taintbench}.
 
The OWASP WebGoat is a deliberately insecure application aiming to teach developers about relevant security vulnerabilities. As a Java Spring application\footnote{https://spring.io/}, it is popular in the community and has been used for evaluating static analyses~\citep{anastatiosFrameworks}. We used this application to evaluate the applicability of \fluentql on real-world scenario, including specifying taint-flow queries and running our Boomerang-based and FlowDroid-based taint analysis. 

We chose to work with the SQL injection as example since it has the most taint-flows in WebGoat. We documented all 17 SQL injection taint-flows in OWASP WebGoat and use them as ground truth. This was manually done by following the directions of the lessons present in WebGoat and inspecting the source code. 

Next, we specified the sensitive methods which includes 17 sources, 1 sanitizer, 2 required propagators, and 2 sinks. We only needed to create two taint-flow queries to be able to cover all types of taint-flows. The Boomerang-based implementation was able to detect all 17 taint-flows. The official FlowDroid implementation (we used version 2.8) was not able to find any taint-flow in WebGoat. We investigated and found out that FlowDroid defines only the return values of the sources as taints. For all taint-flows in WebGoat, the taints are the parameters of the sources. Hence, we adapted FlowDroid to support this and after doing so, the FlowDroid implementation detected 13 taint-flows. Those that were missed are the types that contain a required propagator which is currently not supported by FlowDroid. 

For the second Java application, PetClinic, we followed the same steps as for OWASP WebGoat. We identified and documented five taint-flows of type hibernate injection and two taint-flows of type cross-site request forgery. In this application, all taint-flows were detected by our implementation with Boomerang and our updated version of FlowDroid. Table~\ref{table:projectsJava} shows summary of the Java applications. 

TaintBench is a collection of real-world Android apps that contain malicious behavior in form of taint-flows. These apps have well documented information about the expected taint-flows and should help analyses writers evaluate their tools in a rigorous and fair way. Table~\ref{table:projectsAndroid} summarizes the findings of running our \fluentql implementation with Boomerang as well as with FlowDroid. Out of 25 expected taint-flows among all apps the Boomerang-based implementation found 18 whereas FlowDroid-based implementation found 13. We manually inspected those that were not found and identified two causes which are due to the existing solvers and not the inability of \fluentql to express them. The first cause is the inability to analyze taint-flows through different threads in the code. Due to implicit data-flow behavior of the threads, the existing call graph algorithms have limitation in modeling this correctly. This second cause is that the existing data-flow analyses do not analyze the expressions within path constraints. In the case of our experiments, we found that the call of the source method is within the condition of an IF statement, which is not analyzed by Boomerang nor by FlowDroid. 

The runtime values reported in Table~\ref{table:projectsJava} and Table~\ref{table:projectsAndroid} are the average values over ten runs on a system with Intel(R) Core(TM) i7-8565U CPU @ 1.80GHz, 16 GB RAM with Win-10 OS. 

\begin{lstlisting}[caption={Malisious taint-flow through a file in the \textit{dsencrypt} app from TaintBench}, label=list:fileflow, escapechar=?]{Name}
private void loadClass(Context context) {
	...
	try {
		InputStream is = getAssets().open("ds");  ?\label{lstlisting-5-source}?
		int len = is.available();
		byte[] encrypeData = new byte[len];
		is.read(encrypeData, 0, len); ?\label{lstlisting-5-if1}?
		byte[] rawdata = new DesUtils( DesUtils.STRING_KEY).decrypt(encrypeData);?\label{lstlisting-5-if2}?
		FileOutputStream fos = new FileOutputStream(sourcePathName);?\label{lstlisting-5-if3}?
		fos.write(rawdata);
		fos.close();
	} catch (Exception e) {
		e.printStackTrace();
	}
	try {
		Object[] argsObj = new Object[]{sourcePathName, outputPathName, Integer.valueOf(0)};
		DexFile dx = (DexFile) Class.forName( "dalvik.system.DexFile").getMethod("loadDex", new Class[]{String.class, String.class, Integer.TYPE}).invoke(null, argsObj);?\label{lstlisting-5-sink}?
		...
	} catch (Exception e2) {
	}
}
\end{lstlisting}

\paragraph{Detecting taint-flows through files.} The code in Listing~\ref{list:fileflow} shows the \textit{loadClass} method from the app \textit{dsencrypt}  \footnote{https://github.com/TaintBench/dsencrypt\_samp/blob/master/src/main/java/com/kbstar/\\kb/android/star/ProxyApp.java} which contains the malicious taint-flow. It reads an encrypted zip file from asset folder (source in Line~\ref{lstlisting-5-source}), decrypts it and extracts class.dex which contains malicious code (intermediate statements in the trace are lines \ref{lstlisting-5-if1}, \ref{lstlisting-5-if2}, and \ref{lstlisting-5-if3}). The malicious code is called via reflection (sink in Line~\ref{lstlisting-5-sink}). As reported in the work by Luo et al.\citep{taintbench}, these kind of taint-flows going through files, databases, etc., can not be detected by the existing Android taint analysis tools. With \fluentql, we are now able to model and detect these taint-flows using the and() operator. 

\begin{table}[H]
	\caption{Overview of the evaluated Java projects. Flows/B/F is number of expected taint-flows (vulnerability instances) and those found by Boomerang and FlowDroid, CWE is number of common weakness enumerations (vulnerability types), Runtime is average over ten runs.}
	\begin{tabular}{|l|c|c|c|c|}
		\hline
		\multicolumn{1}{|c|}{\textbf{Project}} & \textbf{\#Classes} & \textbf{\#Flows/B/F}  & \textbf{\#CWE} & \textbf{\#Runtime(s) B/F} \\ \hline
		Catalog                                &        36            &        27/27/25        & 11             &     52.8/43.7                  \\ \hline
		PetClinic                              &         42           &       4/4/4          & 1           &    10.9/14.4                   \\ \hline
		WebGoat                          &         35           &       17/17/13                   &  1             &   30.3/36.7                    \\ \hline
	\end{tabular}
	\label{table:projectsJava}
\end{table}

\begin{table}[H]
	\caption{Overview of the evaluated Android apps from TaintBench. Flows/B/F is number of expected taint-flows (vulnerability instances) and those found by Boomerang and FlowDroid, Runtime is average over ten runs.}
	\begin{tabular}{|l|c|c|c|}
		\hline
		\multicolumn{1}{|c|}{\textbf{App}} & \textbf{\#Classes} & \textbf{\#Flows/B/F}  & \textbf{\#Runtime(s) B/F} \\ \hline
		blackfish                          &         338           &       13/11/11                    &   18.6/29.8                   \\ \hline
		beita\_com\_beita\_contact	&         379                        &        3/1/1                   &   11.2/25.5                    \\ \hline
		phospy                          &         236                        &        2/2/0                   &   8.6/11.5                    \\ \hline
		repane                          &         5                        &        1/1/0                   &   3/4.8                    \\ \hline
		dsencrypt                          &         4           &       1/1/1                &   10.2/4.9                    \\ \hline
		fakeappstore                          &         402                &        3/2/0                  &   23/16.5                    \\ \hline
		fakemart                          &         868                 &        2/0/0             &   27.3/34.9                    \\ \hline
	\end{tabular}
	\label{table:projectsAndroid}
\end{table}

\roundbox{Our Boomerang-based implementation of \fluentql is able to detect all expected taint-flows in the Java Spring applications: OWASP WebGoat and the PetClinic. Among seven real-world Android apps with malicious taint-flows, \fluentql can detect 18 out of 25 expected taint-flows. Those that can not be detected are complex modeling of threads and not considering path conditions.}

\subsection{Threats to Validity}\label{ssec-4-validity}

We next discuss the most relevant threats to the validity of our study design and evaluation based on the threat types by Cook et al.~\citep{threats}.

\paragraph{External validity.} The participation in the study was voluntary. We asked our contacts in industry to invite their software developers. The invitation mentioned that the study would try to compare two domain-specific languages for static analysis. Having this information, it is more likely that the participants have some interest in the design of programming languages and/or static analysis. Hence, there is a threat of having a subject not representative for the entire population of software developers.

\paragraph{Internal validity.} Apart from professional software developers, we invited researchers and master-level students from the university. Previous work has shown that graduate students are valid proxies for software developers in such studies~\citep{developersmet3, developersmet4, developersmet2, developersmet1}. Also, our results confirm that there is no significant correlation between the position of the participant and the performance in the task, thus also confirming that---for the purpose of such studies---researchers and master-level students have coding knowledge comparable to professional developers.

Moreover, the format of within-subjects study design has its own limitations. As both tasks were the same, but for a different DSL and context (vulnerability example), when solving the second task, participants may have be influenced by the first task, known as carryover effects. To deal with this we applied randomization of the order of the tasks. 

\paragraph{Construct validity.} Another threat to validity is the fairness of the tasks. Both DSLs are not equally expressible. This means one may need more or less time to learn a new DSL. To address this, we took into consideration the following points. First, we used vulnerabilities that have the same taint-flow pattern. The Java code shown as an example for each task had the same complexity. Second, for each task, we provided a stub code for the solution. In the case of \codeql, which is more expressible DSL than \fluentql, and has support not only for taint analysis but other analyses too, we asked the participants to focus only on the taint analysis module. Finally, we had three test sessions to adjust the tasks and define what exact information the participants will need to be able to solve each task in under ten minutes. Similarly, a possible threat to validity comes from the design of our study to use the open redirect vulnerability with \fluentql and XSS for \codeql for all participants without switching among half of the participants. To mitigate the threat, we have selected the code examples used in the tasks to have the same structure, i.e. the taint was in both cases created by a call to an HTTP request object and only the sink method differs for each vulnerability. Additionally, while explaining each task, we also explained the vulnerability. While the participants were performing the task, we encouraged them to verbally share their thoughts. After processing the recorded material, we find none of the participants to struggle with understanding the vulnerability itself. 


\section{Related Work}\label{sec-5-relatedwork}

With few exceptions, such as DroidSafe~\citep{droidsafe}, which has hard-coded \srm, the \srm of the existing static analyses can be customized by the user, to some extend.
In this section, we discuss the related approaches summarized in Table~\ref{table:related} that shows design principals of each approach and level of fulfillment to the requirements from Section~\ref{sec-2-motivation}. 

\begin{table}[H]
	\small
	\centering
	\caption{List of related approaches, their design characteristic, and their support of the requirements in Section~\ref{sec-2-motivation}.  \protect\pie{360} - fullfilled, \protect\pie{180} - partially fullfilled, \protect\pie{0} - not fullfilled}
	\setlength\tabcolsep{.5mm}
	\begin{tabular}{c|l|c|c|c|c|c|c|cccccc}
		&Approach & \rotatebox{90}{declarative} & \rotatebox{90}{general-like} & \rotatebox{90}{object-oriented} & \rotatebox{90}{constraints} &  \rotatebox{90}{pattern-based}  & \rotatebox{90}{annotations}& \requirement{1} & \requirement{2} & \requirement{3} & \requirement{4} & \requirement{5} & \requirement{6} \\ \hline
		
		\multirow{4}{*}{Graph-based} & CxQL \citep{checkmarx} & & X & X & &  & & \pie{360} & \pie{180} & \pie{180} & \pie{360} & \pie{360} & \pie{0} \\ 
		&\codeql \citep{lgtm} & X & X & X  & &  & &  \pie{360} & \pie{180} & \pie{180} & \pie{360} & \pie{360} & \pie{0} \\
		&PIDGIN \citep{pidgin} & & & X & X & & & \pie{360} & \pie{360} & \pie{180} & \pie{0} & \pie{0} & \pie{0} \\
		&IncA \citep{inca} & & & & X & X &  & \pie{180} & \pie{180} & \pie{0} & \pie{0} & \pie{180} & \pie{0} \\ \cline{1-1}
		
		\multirow{2}{*}{Typestate}&CrySL \citep{crySL}  & X & & & & X  & & \pie{360} & \pie{360} & \pie{180} & \pie{0} & \pie{180} & \pie{0} \\
		&PQL \citep{pql} & X & & & X & & & \pie{360} & \pie{360} & \pie{180} & \pie{0} &  \pie{0} & \pie{0} \\ \cline{1-1}
		
		\multirow{6}{*}{Other}&CheckersF. \citep{checkersFramework} & & & & & & X &\pie{180} & \pie{360}  & \pie{0} & \pie{360} & \pie{0} & \pie{0} \\	
		&Apposcopy \citep{Apposcopy}  & X & & & & & X & \pie{180} & \pie{360} & \pie{180} & \pie{0} & \pie{0} & \pie{180} \\
		&Athena \citep{leDSL} & X & & & X & X &  & \pie{360} & \pie{360} & \pie{0}& \pie{0} & \pie{0} & \pie{0} \\
		&AQL \citep{aql} & & & & X & X & & \pie{180} & \pie{360} & \pie{0} & \pie{0} & \pie{360} & \pie{0} \\
		&WAFL \citep{f4f} & & X & X & &  & & \pie{180} & \pie{180} & \pie{180} & \pie{0} & \pie{0} & \pie{0} \\
		&Saluki \citep{saluki} & X & & & X & X & & \pie{180} & \pie{360} & \pie{180} & \pie{0} & \pie{0} & \pie{0}\\
		\hline
		&\fluentql & & & & & X & & \pie{360} & \pie{360} & \pie{360} & \pie{360} & \pie{360} & \pie{360}\\
		\hline
	\end{tabular}
	\label{table:related}
\end{table}

\subsection{Graph-based approaches} 
We group DSLs in this category that allow users to explicitly manipulate graph structures to specify code patterns. 

CxQL, the DSL of the tool Checkmarx~\citep{checkmarx}, is general-purpose-like and object-oriented language. It supports a wide range of \srm (\requirement{1}). The flow propagation is done implicitly via the graph patterns (\requirement{2}). As a commercial tool it provides well integrated workflow for the users (\requirement{4} and \requirement{5}). Since CxQL is capable of expressing a broad range of graph properties, it is hard to integrate it to a generic taint analysis (\requirement{6}) as it is bound to the tool's core analysis. 

\codeql has good integration in the developers' workflow (\requirement{5}) with plugins for popular IDEs as well as web-based interface. It is a declarative language with support for predicates and object-oriented design. The DSL can express a wide range of properties similar to CxQL. 

PIDGIN \citep{pidgin} follows the object-oriented design. It is not designed for taint analysis, therefore it is hard to integrate it in a generic way (\requirement{6}). PIDGIN does not provide tool integrations for software developers. 

IncA \citep{inca} is a DSL for specification of the rules for incremental execution of static analyses. Compared to the other DSL in this category it is the least expressive and targets very specific domain. The \srm and flow propagation can be expressed through the graph patterns (\requirement{1}, \requirement{2}). User-defined messages are not supported (\requirement{4}).

\subsection{Typestate approaches.} 
This category consists of two DSLs, i.e. CrySL and PQL, which are designed for typestate analysis, that detects the incorrect API usage.

CrySL \citep{crySL} enables cryptography experts to specify the correct usage of the crypto API making it not suitable for generic taint analysis (\requirement{6}). The DSL is declarative with mechanism based on predicates and constraints. It has a full support for \srm and flow propagation (\requirement{1}, \requirement{2}). The tool support is maintained. 

PQL \citep{pql} is declarative DSL with specifications comparable to CrySL. Compared to \fluentql, the PQL's syntax significantly differs from regular Java syntax making it difficult to use for non-experts. PQL does not ship with available tooling support for the users (\requirement{6}). 

\subsection{Other approaches.} 

The approach used in CheckersFramework~\citep{checkersFramework} is based on the annotations \textit{@tainted} and \textit{@untainted} which developers can use to annotate their code to mark custom \srm. The annotation specification requires additional manual work, which - in the case of legacy code - is even infeasible (\requirement{5}). The CheckersFramework allows to configure the messages that are reported (\requirement{4}). 

Apposcopy \citep{Apposcopy} is a Android-specific taint analysis (\requirement{6}) with a Datalog-based DSL for data-flow and control-flow predicates. The sources, sinks, and propagators are specified in form of annotations (\requirement{1}, \requirement{2}). We were not able to find tooling support (\requirement{5}) for the language. 

Athena \citep{leDSL} is a declarative DSL based on patterns and constraints with explicit support for \srm and flow propagations (\requirement{1}, \requirement{2}). Athena does not support user-defined messages and is tightly coupled to a generator of analysis configurations. 

AQL \citep{aql} is an Android-specific (\requirement{6}) querying language for taint flow results from different taint analysis tools. It supports specifications of sources and sinks (\requirement{1}) and flow propagations (\requirement{2}) and provides workflow with tool support for developers to query taint results from multiple tools (\requirement{5}). 

WAFL is the DSL of the F4F approach \citep{f4f} and is a general-purpose-like language with object-oriented design that allows the specification of reflective behavior in frameworks so that static analyses can propagate the flow. WAFL has partial support for \srm and flow propagation (\requirement{1}, \requirement{2}). Its main purpose is modeling of frameworks.  

Finally, Saluki~\citep{saluki} is a declarative DSL where method patterns can be specified. \srm and flow propagation are supported (\requirement{1}, \requirement{2}), but complex flows not (\requirement{3}). 


\section{Conclusion}\label{sec-6-conclusion}

Static and dynamic taint analyses can detect many popular vulnerability types. 
Using them during development time can reduce the costs.
To use taint analysis tools efficiently, software developers need to configure them to their contexts. 

We proposed \fluentql, a new domain-specific language for taint analysis designed to be used by software developers. 
\fluentql is able to express many taint-style vulnerability types. 
It supports single-step as well as multi-step taint-flow queries. 
Moreover, taint-flow queries can be combined with \emph{and} operator to express parallel taint-flows. \fluentql uses Java fluent interface as a concrete syntax. 
Its semantics is independent of a concrete implementation of taint analysis, making it easy for integration into existing tools. 

In a comparative, within-subjects user study, \fluentql showed to be more usable for software developers than \codeql, a state-of-the-art DSL of the commercial tool LGTM. Participants were faster in solving the task of specifying a taint-flow query for given vulnerability type in \fluentql than in \codeql. 

In future, we plan to make \fluentql more expressive. With \fluentql, only method calls can be modeled as sources and sinks. We will add support for modeling specific variables, such as constant values for detecting hardcoded credentials and nullable variables for detecting null pointer dereferences. Moreover, we will add support for regular expressions to ease the specification of methods in case of method overloading in Java. Additionally, we plan on investigating the usefulness of incorporating the concept of entry point as part of \fluentql. Finally, we are working on further evaluations of the applicability of \fluentql in other real-world large-scale Java applications and creating new queries for detecting more security vulnerabilities.  

\begin{acknowledgements}
	We gratefully acknowlede the funding by the project "AppSecure.nrw - Security-by-Design of Java-based Applications" of the European Regional Development Fund (ERDF-0801379). We thank Ranjith Krishnamurthy and Abdul Rehman Tareen for their contribution to the implementation of the MagpieBridge server.
\end{acknowledgements}

\textbf{Authors' contributions}: The first author is the main contributor to this research. The second and fourth author contributed with conceptual ideas and feedback. The third author contributed to the implementation and with ideas for the concrete syntax of fluentTQL.

\textbf{Data availability}: https://fluenttql.github.io/  

\textbf{Code availability}: https://github.com/secure-software-engineering/secucheck

\section*{Declarations}

\textbf{Ethics approval}: The user study design has been approved for ethical correctness by one of the companies participated in the study as well as by the corresponding head of department at Fraunhofer IEM. 

\textbf{Consent to participate and publication}: For the user study described in section 4, all 26 participants signed a written consent form in which they agreed to participate voluntarily in the study. They also agreed that the collected data can be used for research publication.  The written consent form was obtained from all participants before the study. 

\textbf{Conflicts of interest/Competing interests}: Not applicable

\bibliographystyle{spbasic}
\bibliography{references}

\end{document}